\documentclass[preprint,12pt]{elsarticle}

\usepackage{amssymb}
\usepackage{amsmath}
\usepackage{hyperref}
\hypersetup{colorlinks,allcolors=black}
\usepackage{pbox}
\usepackage{graphicx}
\usepackage{caption}
\usepackage{multicol}
\usepackage{multirow}
\usepackage{rotating}
\usepackage{float}

\usepackage{booktabs}
\usepackage{xcolor}
\usepackage{colortbl}
\usepackage{array}
\usepackage{subfig}

\definecolor{headercolor}{RGB}{220,220,220}
\definecolor{rowcolor1}{RGB}{245,245,245}
\definecolor{rowcolor2}{RGB}{255,255,255}
\definecolor{lightgray}{RGB}{245,245,245}

\journal{}

\begin{document}

\begin{frontmatter}

\title{Benchmarking and Explaining Deep Learning Cortical Lesion MRI Segmentation in Multiple Sclerosis} 

\author[unil,chuv,hes,cibm]{Nataliia Molchanova\corref{cor1}}
\cortext[cor1]{Corresponding author.}
\ead{nataliia[dot]molchanova[at]unil[dot]ch}

\author[unibas1,unibas2,unibas3,genova]{Alessandro Cagol}
\author[unibas1,unibas2,unibas3]{Mario Ocampo--Pineda}
\author[unibas1,unibas2,unibas3]{Po--Jui Lu}
\author[unibas1,unibas2,unibas3,unibas4]{Matthias Weigel}
\author[unibas1,unibas2,unibas3]{Xinjie Chen}

\author[sinai]{Erin S. Beck}

\author[nih]{Charidimos Tsagkas}
\author[nih]{Daniel S. Reich}

\author[uclouvain]{Colin Vanden Bulcke}
\author[uclouvain]{Anna Stölting}
\author[uclouvain]{Serena Borrelli}
\author[uclouvain]{Pietro Maggi}


\author[hes]{Adrien Depeursinge}

\author[unibas1,unibas2,unibas3]{Cristina Granziera}

\author[hes,unige]{Henning Müller}

\author[cibm,chuv,unil]{Pedro M. Gordaliza\corref{label2}}
\author[cibm,chuv,unil]{Meritxell Bach Cuadra\corref{label2}}
\cortext[label2]{Equal contributions.}

\affiliation[unil]{
organization={Faculty of Biology and Medicine, University of Lausanne (UNIL)}, 
city={Lausanne},
country={Switzerland}
}
\affiliation[chuv]{
organization={Radiology Department, Lausanne University Hospital (CHUV)}, 
city={Lausanne},
country={Switzerland}
}
\affiliation[hes]{
organization={MedGIFT, Institute of Informatics, School of Management, HES--SO Valais--Wallis University of Applied Sciences and Arts Western Switzerland},
city={Sierre},
country={Switzerland}
}
\affiliation[cibm]{
organization={CIBM Center for Biomedical Imaging},
city={Lausanne},
country={Switzerland}
}
\affiliation[unige]{
organization={Department of Radiology and Medical Informatics, University of Geneva},
city={Geneva},
country={Switzerland}
}
\affiliation[unibas1]{
organization={Translational Imaging in Neurology (ThINK) Basel, Department of Medicine and Biomedical Engineering, University Hospital Basel and University of Basel},
city={Basel},
country={Switzerland}
}
\affiliation[unibas2]{
organization={Multiple Sclerosis Center, Department of Neurology, University Hospital Basel},
city={Basel},
country={Switzerland}
}
\affiliation[unibas3]{
organization={Research Center for Clinical Neuroimmunology and Neuroscience Basel (RC2NB), University Hospital Basel and University of Basel},
city={Basel},
country={Switzerland}
}
\affiliation[unibas4]{
organization={Division of Radiological Physics, Department of Radiology, University Hospital Basel},
city={Basel},
country={Switzerland}
}
\affiliation[genova]{
organization={ Dipartimento di Scienze della Salute, Università degli Studi di Genova},
city={Genova},
country={Italy}
}
\affiliation[sinai]{
organization={Department of Neurology, Icahn School of Medicine at Mount Sinai}, 
city={New York City},
country={USA}
}
\affiliation[nih]{
organization={Translational Neuroradiology Section, National Institute of Neurological Disorders and Stroke, National Institutes of Health},
city={Bethesda},
country={USA}
}
\affiliation[uclouvain]{
organization={Neuroinflammation Imaging Lab (NIL), Université catholique de Louvain},
city={Brussels},
country={Belgium}
}

\begin{abstract}
Cortical lesions (CLs) have emerged as valuable biomarkers in multiple sclerosis (MS), offering high diagnostic specificity and prognostic relevance. However, their routine clinical integration remains limited due to subtle magnetic resonance imaging (MRI) appearance, challenges in expert annotation, and a lack of standardized automated methods. We propose a comprehensive multi-centric benchmark of CL detection and segmentation in MRI. A total of 656 MRI scans, including clinical trial and research data from four institutions, were acquired at 3T and 7T using MP2RAGE and MPRAGE sequences with expert-consensus annotations. We rely on the self-configuring nnU-Net framework, designed for medical imaging segmentation, and propose adaptations tailored to the improved CL detection. We evaluated model generalization through out-of-distribution testing, demonstrating strong lesion detection capabilities with an F1-score of 0.64 and 0.5 in and out of the domain, respectively.
We also analyze internal model features and model errors for a better understanding of AI decision-making. Our study examines how data variability, lesion ambiguity, and protocol differences impact model performance, offering future recommendations to address these barriers to clinical adoption.
To reinforce the reproducibility, the implementation and models will be publicly accessible and ready to use at \href{https://github.com/Medical-Image-Analysis-Laboratory/}{GitHub}  at \href{https://doi.org/10.5281/zenodo.15911797}{Zenodo}.

\end{abstract}

\begin{graphicalabstract}
\end{graphicalabstract}

\begin{highlights}
\item nnU-Net benchmarks robust cortical lesion segmentation in large multi-center dataset.

\item Comparative study reveals key nnU-Net modifications enhancing CL segmentation accuracy.

\item Expert review validates clinical relevance of predicted false-positive cortical lesions.

\item Bottleneck feature analysis explores U-Net generalization limits across clinical centers.

\item Study highlights practical barriers hindering clinical integration of CL segmentation.
\end{highlights}

\begin{keyword}

Multiple sclerosis \sep Cortical lesions  \sep Segmentation \sep Detection \sep Magnetic Resonance Imaging \sep Brain \sep Deep learning \sep Trustworthy AI

\end{keyword}

\end{frontmatter}

\section{Introduction}
\label{intro}

\begin{figure}[h!]
    \centering
    \includegraphics[width=0.6\linewidth]{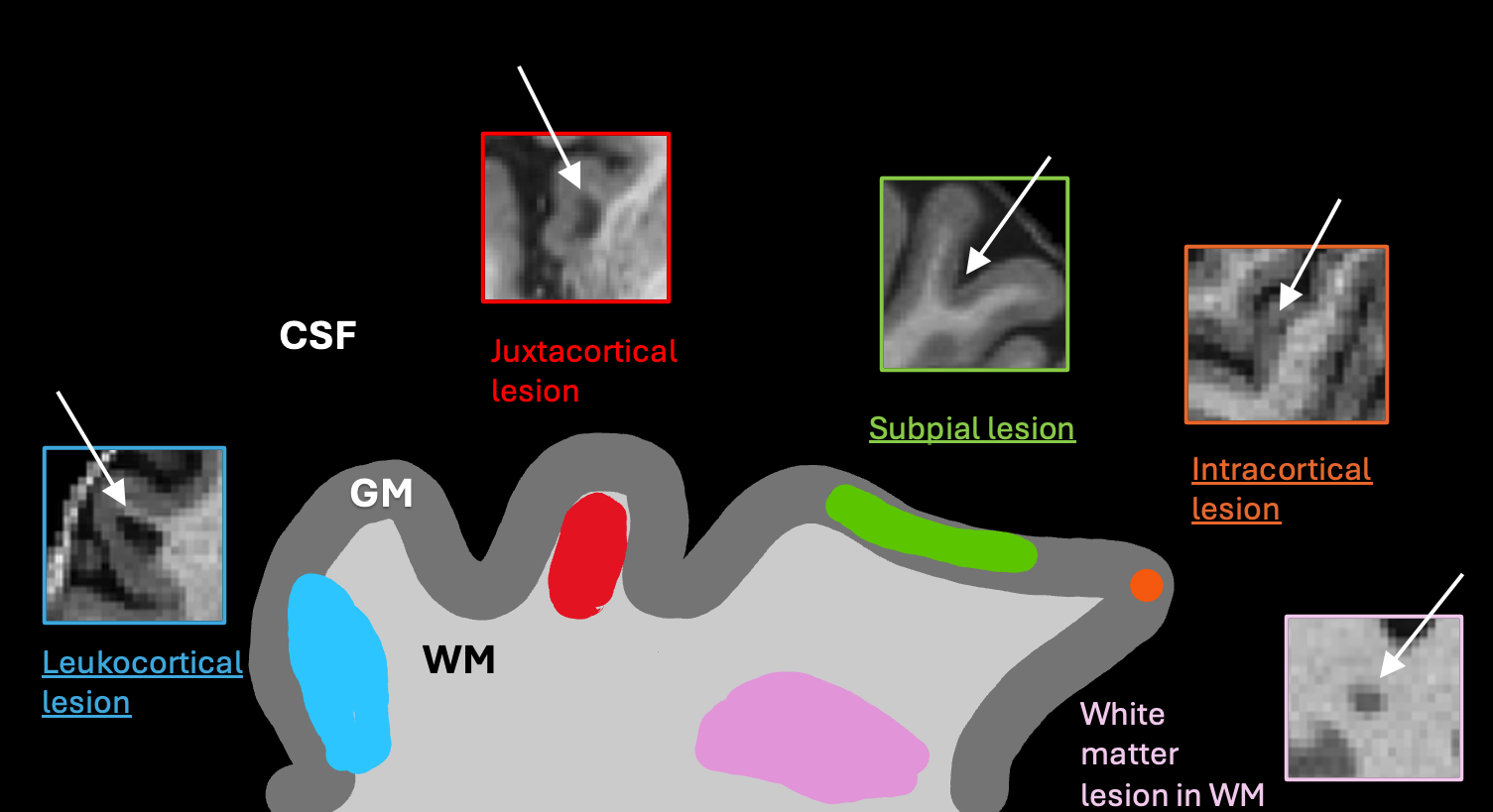}
    \caption{Types of MS lesions characterized through their location with respect to the cortex. Lesions appear on MP2RAGE MRI as hypointense regions (white arrow pointers). CL types are \underline{underlined}, the rest are WMLs. Leukocortical (CL) and juxtacortical (WML) are sometimes difficult to distinguish due to unclear cortical involvement. CSF - cerebrospinal fluid, GM - gray matter, WM - white matter. }
    \label{fig:lestype}
\end{figure}

Multiple sclerosis (MS) is a chronic inflammatory and neurodegenerative disease of the central nervous system, affecting over 2 million people worldwide \cite{reich_multiple_2018, portaccio_multiple_2024}. Magnetic resonance imaging (MRI) is central to the diagnosis and monitoring of MS, enabling the detection of characteristic brain lesions~\cite{hemond_magnetic_2018, filippi_association_2019, wattjes_2021_2021, kolb_pathology_2022}. White matter lesions (WMLs) have served as the primary radiological hallmark, guiding diagnosis and treatment decisions \cite{hemond_magnetic_2018}. In recent years, additional imaging biomarkers—such as cortical lesions (CLs)\cite{beck_cortical_2022, cagol_diagnostic_2024}, paramagnetic rim lesions~\cite{absinta_association_2019, maggi_paramagnetic_2020}, and the central vein sign~\cite{sati_central_2016, cagol_diagnostic_2024} — have gained recognition for their ability to support the diagnosis or to capture disease activity more comprehensively. 
CLs are of clinical interest due to their association with cognitive impairment, progressive disability, and early-stage onset \cite{beck_cortical_2022, beck_contribution_2024, cagol_diagnostic_2024}. Indeed, CLs represent diagnostic values and were incorporated in the 2017 revision of the \textit{McDonald criteria} due to their high specificity for MS \cite{thompson_diagnosis_2018}. Unlike WMLs that appear in numerous neuroinflammatory and neurovascular conditions, the presence of a CLs enhances diagnostic certainty for MS and helps differentiate it from clinical conditions mimicking MS \cite{thompson_diagnosis_2018, cagol_diagnostic_2024}. This distinction is crucial for preventing misdiagnosis and inappropriate treatment, particularly as the McDonald criteria generally favor sensitivity over specificity. Yet, despite their clinical relevance, CLs (as distinct from juxtacortical lesions) remain underused in routine practice due to several technical barriers: their small size (particularly for small intracortical lesions), low contrast on standard MRI sequences, and the need for expert radiological interpretation \cite{kolb_pathology_2022}. Different MS lesion types are illustrated in \autoref{fig:lestype}.

Automated segmentation methods offer a promising avenue to facilitate the adoption of CL assessment into clinical workflows. They can reduce the burden of manual annotation, increase reproducibility, and enable broader access to quantitative lesion metrics. However, the development and deployment of such tools are hindered by several caveats: limited annotated datasets, scanner and protocol variability, and poor generalization across sites. CL segmentation, in particular, is further challenged by \textit{high class imbalance} and substantial heterogeneity in lesion appearance. These limitations have slowed the translation of automated CL assessment tools into clinical use.

Deep learning (DL) approaches for CL segmentation have been limited by the scarcity of adequately annotated training data.
Consequently, initial DL approaches employed joint segmentation strategies, where CLs and WMLs were treated as a single lesion class without distinction. Concretely, in \cite{rosa_multiple_2020}, authors proposed a shallow 3D U-Net model~\cite{ronneberger_u-net_2015, cicek_3d_2016} for this joint segmentation using fluid-attenuated inversion recovery (FLAIR) and magnetization prepared 2 rapid gradient echo (MP2RAGE)~\cite{kober_mp2rage_2012} sequences.  While this approach reported improved detection compared to previous non-DL methods \cite{fartaria_longitudinal_2019}, the joint classification inherently inflated CL detection performance, as predictions in juxtacortical white matter regions—which would constitute false positives in CL-specific segmentation—were correctly classified as WMLs, thereby avoiding penalization despite anatomical imprecision.

Dedicated CL segmentation with DL was first explored using ultra-high-field (7T) MRI, which enhances the visibility of CLs~\cite{beck_improved_2018}.  The CLAIMS framework \cite{la_rosa_multiple_2022} demonstrated substantial improvements in CL-specific detection, particularly for leukocortical and subpial lesions. Crucially, this approach required explicit CL identification, meaning any detections in white matter regions (including juxtacortical areas) were counted as false positives rather than alternative lesion types. When accounting for this methodological difference, the CL detection performance was comparable to previous joint segmentation approaches, but with significantly improved specificity for cortical tissue. Despite its effectiveness, the application remains limited due to the low availability of 7T scanners in typical clinical settings and modest inter-rater reliability even at these high fields.

Further advances aimed at leveraging enhanced contrast visibility on conventional scanners include combining multiple advanced contrasts or synthetic sequences. Reference \cite{gordaliza_fluid_2025} investigated the impact of fluid and white matter suppression (FLAWS) alongside MP2RAGE using the 3D nnU-Net~\cite{isensee_nnu-net_2021} framework. This CL-specific approach at 3T achieved detection results comparable to earlier methods while maintaining the anatomical precision required for clinical CL assessment. However, its clinical translation is hindered by the requirement for FLAWS sequences not routinely implemented in clinical protocols, potentially limiting widespread adoption. Similarly, \cite{dwyer_quantifying_2025} evaluated retrospective segmentation methods leveraging legacy clinical trial data, introducing novel contrast combinations such as FLAIR$^2$, T1/T2 ratio, and AI-derived synthetic double inversion recovery (DIR) contrasts. Their findings highlighted improved CL detection when using synthetic MRI-derived contrast maps and emphasized the persistent challenges related to variability across scanner platforms and clinical settings.  

While these studies demonstrate significant progress in automated CLs segmentation, a comprehensive evaluation of different DL architectures and their comparative performance remains unexplored. Most works have focused on optimal input contrast combinations, leaving questions about model architecture selection, generalizability across clinical settings, and failure case analysis largely unaddressed. 

Furthermore, clinical deployment of DL models requires robust performance across diverse imaging protocols, scanner vendors, and field strengths that differ significantly from controlled research environments. Previous generalizability assessments have been limited in scope, with studies testing on relatively small out-of-domain cohorts (36 subjects in \cite{rosa_multiple_2020}, 20 subjects in \cite{la_rosa_multiple_2022}) using similar scanner types and protocols. Such evaluations may not adequately reflect the heterogeneity encountered in real-world clinical settings, where scanner vendors, acquisition parameters, and patient populations vary substantially. Additionally, the complex nature of CL detection, where subtle anatomical distinctions determine diagnostic relevance, necessitates comprehensive error analysis to understand model behavior and failure modes—an aspect that has not been systematically addressed in prior CL segmentation literature.

\section{Contributions}
We present a comprehensive benchmark for CL segmentation using a large multi-center dataset comprising 656 scans from 4 medical centers across different field strengths (3T/7T) and acquisition protocols. Our framework leverages commonly acquired MP2RAGE and MPRAGE sequences to ensure clinical applicability. Specifically, our contributions include:
\begin{itemize}

    \item A systematic architectural comparison establishing optimal model configurations for CL segmentation across diverse clinical settings,

    \item  Comprehensive out-of-distribution evaluation on 224 subjects across different scanner vendors, acquisition protocols, and patient cohorts—the largest and most heterogeneous OOD assessment in CL segmentation to date,
    \item Systematic error analysis in CL segmentation, identifying failure modes and providing actionable insights for model improvement and clinical interpretation,
    \item Public release of optimized model weights and implementation framework to enable widespread in adoption in clinical and research environments.
\end{itemize}

\section{Materials and methods}

\subsection{Data}

Four medical centers shared their longitudinal and cross-sectional data for this study, including A) Basel University Hospital (USB), Switzerland, B) University Hospital of Lausanne (CHUV), Switzerland, C) National Institutes of Health, USA, and D) Catholic University of Louvain (UCLouvain), Belgium. The data information is provided in Table \ref{tab:data}.

\subsubsection{Manual annotation}
The scans were manually annotated using local resources and annotation guidelines. In sites A-D, all lesions involving the cortex were included in the CL masks, including intracortical, leukocortical, or subpial, typically extending for at least 3mm in the longest diameter. In A, CLs were segmented on 1mm isotropic MP2RAGE through an independent assessment of two raters (neurologist and MD with 5 years of neuroimaging experience). In B, CLs were detected using $1\times1\times1.2$mm$^3$ 3D FLAIR, 3D DIR, and MP2RAGE acquired at 3T by 2 independent raters (an experienced neurologist and radiologist), followed by a consensus review; then, a trained technician manually delineated the consensus lesions. In C, 7T visible CLs were segmented on 0.5mm isotropic MP2RAGE (median of 4 acquisitions in the same scanning sessions) and 0.5mm multi-echo T2*w GRE by 2 independent raters (neurologists with 2 and >10 years of MS neuroimaging experience), followed by a consensus review. In C, 3T CLs were independently segmented on 1mm isotropic 3D FLAIR and MP2RAGE by 2 raters (neurologists with 2 and 4 years of MS neuroimaging experience), followed by a consensus review. In D, the detection of CLs was performed using 0.7mm isotropic 3D DIR and MP2RAGE by 3 independent raters (a neuroscientist and two neurologists with 3, 5, and 10 years of MS neuroimaging experience, respectively), followed by a consensus review; the trained neuroscientist rater performed the manual delineation of consensus lesions.

All scans from site B and 163 scans from site A (the first time point) had information about the lesion types (either from MP2RAGE or FLAIR) and included FLAIR-based WML segmentation masks. Subjects from site C 7T data had information about the lesion types and segmentation of \textit{potential lesions}. Potential lesions were the ones where expert raters could not conclude whether lesions were CL or not, and were not used during training or evaluation. The rest of the test data did not contain WML masks or lesion phenotyping. 

\clearpage

\begin{sidewaystable}[h!]
\caption{Data description. RR - relapsing-remitting, PP - primary-progressive, SP - secondary-progressive, CIS - clinically isolated syndrome, Q2 - second quartile, IQR - interquartile range.}
\scriptsize
\centering
    \begin{tabular}{*{5}{p{4cm}}} 
    \toprule
    \cellcolor{headercolor}\textbf{Medical center} & \cellcolor{headercolor}\textbf{A} & \cellcolor{headercolor}\textbf{B} & \cellcolor{headercolor}\textbf{C} & \cellcolor{headercolor}\textbf{D} \\
    \midrule
    \rowcolor{headercolor}\multicolumn{5}{c}{\textbf{Demographics}} \\
    \midrule
    \rowcolor{lightgray}\textbf{\# patients} & 163 & 43 & 35 & 112 \\
    \rowcolor{white}\textbf{Phenotype} & RR (97), PP (22), SP (44) & RR & RR (24), PP (1), SP (9) & CIS (3), RR (44), PP (8), SP (17), MS-mimic (40) \\
    \rowcolor{lightgray}\textbf{\% female} & 60\% & 58\% & 63\% & 70\% \\
    \rowcolor{white}\textbf{EDSS Q2 (IQR)} & 3 (1.5-4.5) & 1.5 (1.5-4) & 2.5 (1-4) & 2.5 (2-4.5) \\
    \rowcolor{lightgray}\textbf{Timepoints} & 2 & 2 (2 year distance) & 3 & 1 \\
    \midrule
    \rowcolor{headercolor}\multicolumn{5}{c}{\textbf{MRI}} \\
    \midrule
    \rowcolor{lightgray}\textbf{Scanner} & Magnetom Prisma, Siemens Healthineers & Magnetom Trio, Siemens Healthineers & 3T Skyra and 7T whole-body research system, Siemens Healthcare & SIGNA\texttrademark{}, GE Health Care \\
    \rowcolor{white}\textbf{Field strength} & 3T & 3T & 3T, 7T & 3T \\
    \rowcolor{lightgray}\textbf{MPRAGE resolution, acquisition parameters (TR/TE/TI, ms)} & - & $1.0\times1.0\times1.2$, 2300/2.84/900 & - & $1\times1\times1$, 2186/3/900 \\
    \rowcolor{white}\textbf{MP2RAGE resolution, acquisition parameters (TR/TE/TI1/TI2, ms)} & $1.0\times1.0\times1.0$, 5000/2.98/700/2500 & $1.0\times1.0\times1.2$, 5000/2.84/700/2500 & for 3T: $0.8\times0.8\times0.8$, 5000/2.9/700/2500, for 7T: $0.5\times0.5\times0.5$, 6000/5/800/2700 & $1\times1\times1$, 5000/3/700/2500 \\
    \midrule
    \rowcolor{headercolor}\multicolumn{5}{c}{\textbf{Processing}} \\
    \midrule
    \rowcolor{lightgray}\textbf{Registration} & - & - & - & Registration to 3D EPI with 0.67x0.67x0.67mm \\
    \rowcolor{white}\textbf{Skull-stripping} & \multicolumn{2}{p{7cm}}{HD-BET~\cite{isensee_automated_2019} to FLAIR + morphological dilation (2 iterations, 3x3x3 structural element)} & SynthSeg~\cite{billot_synthseg_2023} to MP2RAGE + manual correction for 7T & HD-BET to MPRAGE + morphological dilation \\
    \bottomrule
    \end{tabular}
    \label{tab:data}
\end{sidewaystable}

\clearpage

\subsubsection{Data splitting}
The data was split into training (Train), in-domain pure testing (Test-in), and out-of-domain (OOD) testing (Test-out). The data from site D was used for the OOD testing, due to the difference in the scanner manufacturer, compared to the other centers. The medical center possesses the clinically used MPRAGE modality and data for both MS and clinical conditions mimicking MS (MS-mimic) patients; thus, we can test the applicability of the proposed framework to support MS differential diagnosis. The rest of the data A-C were separated on Train and Test-in by using a custom stratified split approach that preserved both subject-level grouping and data distribution characteristics. The method first aggregated data at the subject-site level and then applied stratification techniques while ensuring subjects from the same site remained together in the same split. The implementation maintained similar distributions of site, lesion count, and total lesion volume between splits by creating composite stratification categories. The continuous variables (number of lesions and total volume) were discretized into 5 quantile-based bins, and single-sample stratification groups were merged with their nearest neighbors based on similarity in lesion metrics while respecting site constraints. This approach allocated $80\%$ of the in-domain data to training and $20\%$ to testing, with reproducibility ensured through fixed random seeding. Distribution statistics were verified through a comprehensive comparative analysis that examined the percentage distributions of categorical variables across splits, assessed descriptive statistics (mean, standard deviation, quartiles) of the continuous variables in each partition, and confirmed the preservation of subject-site integrity by counting unique subject-site combinations in the resulting datasets. 

\begin{figure}[h!]
    \centering
    \includegraphics[width=0.73\linewidth]{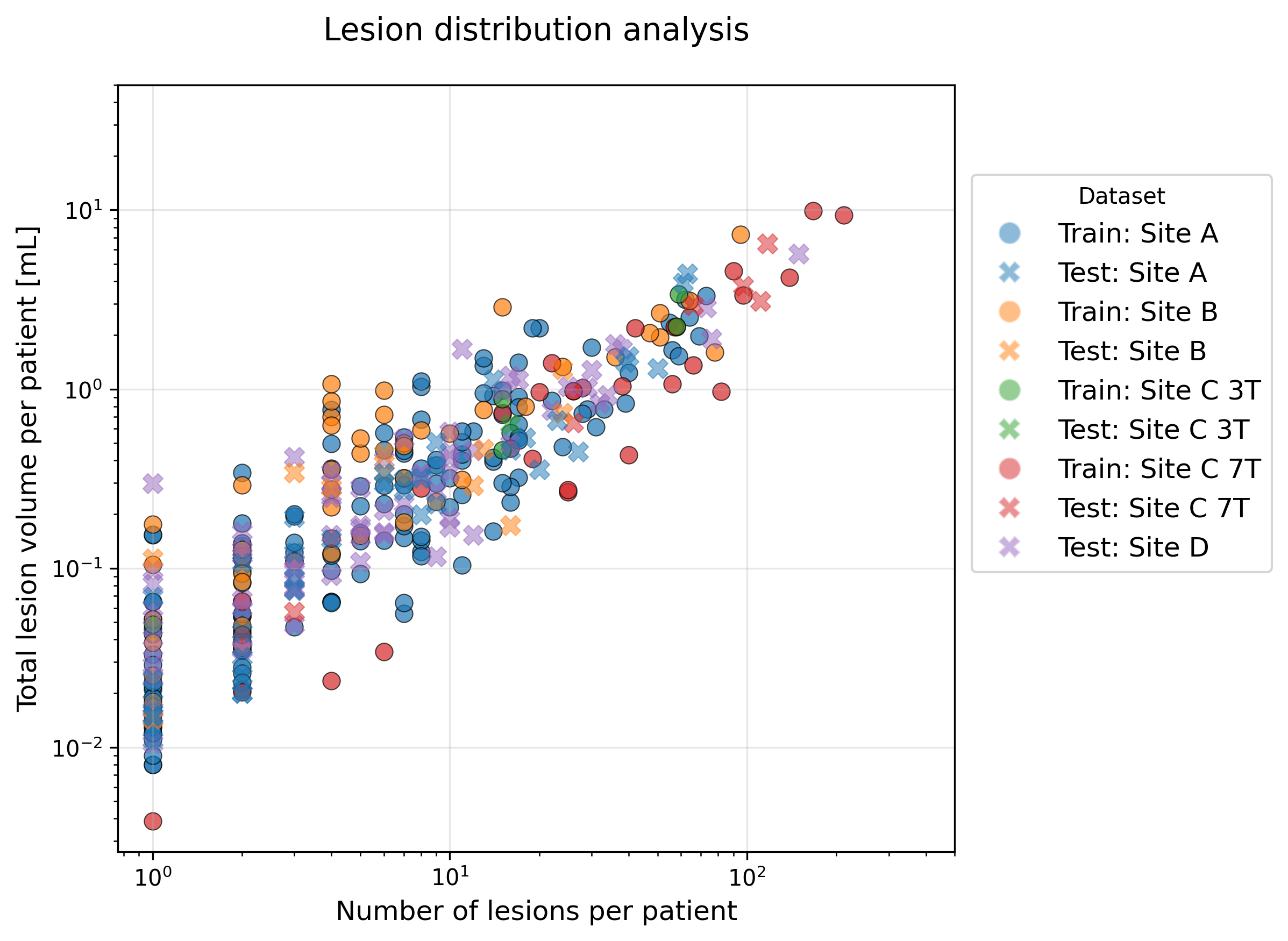}
    \caption{Distribution of total lesion volume in milliliters and number of lesions per patient across different medical sites.}
    \label{fig:lesionload}
\end{figure}

Overall, the differences between A-D sites are driven by varying acquisition, annotation protocols, and patient cohorts. The differences in the distributions of lesions across different sites and domain are shown in Figure \ref{fig:lesionload}. Sites A, B, and C have patients with different total lesion volumes and numbers of lesions per patient; however, this distribution is skewed towards smaller lesion volumes and counts. Site C, containing mostly ultra-high-field 7T data, has higher total lesion volumes and counts per patient since the visibility of CLs on 7T  increased sensitivity for CL detection~\cite{madsen_imaging_2021}. Moreover, the annotations performed on 7T MP2RAGE and T2-star-weighted (T2*w) GRE images resulted in the detection of many subpial lesions. 


\begin{table}[h!]
\caption{Details of the data split. Modality 1 - MPRAGE, 2 - MP2RAGE.} 
\scriptsize
\centering
\begin{tabular}{p{1.6cm}p{0.8cm}p{0.8cm}p{0.8cm}*{4}{p{1.1cm}}}
\toprule
\cellcolor{headercolor}\textbf{Set name} & \multicolumn{3}{c}{\cellcolor{headercolor}\textbf{Partition}} & \multicolumn{4}{c}{\cellcolor{headercolor}\textbf{Site (Number of Scans)}} \\
\cmidrule(lr){2-4}\cmidrule(lr){5-8} 
\cellcolor{headercolor} & \cellcolor{headercolor}\textbf{Moda-lity} & \cellcolor{headercolor}\textbf{7T} & \cellcolor{headercolor}\textbf{MS-mimic} & \cellcolor{headercolor}\textbf{A} & \cellcolor{headercolor}\textbf{B} & \cellcolor{headercolor}\textbf{D} & \cellcolor{headercolor}\textbf{C} \\ 
\midrule
\cellcolor{lightgray}\multirow{-1}{*}{\centering Train}
\cellcolor{lightgray} & \cellcolor{lightgray}2 & \cellcolor{lightgray} & \cellcolor{lightgray} & \cellcolor{lightgray}191 & \cellcolor{lightgray}60 & \cellcolor{lightgray}5 & \cellcolor{lightgray}0 \\
\cellcolor{lightgray} & \cellcolor{lightgray}2 & \cellcolor{lightgray}\checkmark & \cellcolor{lightgray} & \cellcolor{lightgray}0 & \cellcolor{lightgray}0 & \cellcolor{lightgray}31 & \cellcolor{lightgray}0 \\
\cellcolor{lightgray} & \cellcolor{lightgray}1 & \cellcolor{lightgray} & \cellcolor{lightgray} & \cellcolor{lightgray}0 & \cellcolor{lightgray}60 & \cellcolor{lightgray}0 & \cellcolor{lightgray}0 \\
\midrule
\cellcolor{white}\multirow{-1}{*}{\centering Test-in}
\cellcolor{white} & \cellcolor{white}2 & \cellcolor{white} & \cellcolor{white} & \cellcolor{white}50 & \cellcolor{white}14 & \cellcolor{white}1 & \cellcolor{white}0 \\
\cellcolor{white} & \cellcolor{white}2 & \cellcolor{white}\checkmark & \cellcolor{white} & \cellcolor{white}0 & \cellcolor{white}0 & \cellcolor{white}6 & \cellcolor{white}0 \\
\cellcolor{white} & \cellcolor{white}1 & \cellcolor{white} & \cellcolor{white} & \cellcolor{white}0 & \cellcolor{white}14 & \cellcolor{white}0 & \cellcolor{white}0 \\
\midrule
\cellcolor{lightgray}\multirow{-1}{*}{\centering Test-out}
\cellcolor{lightgray} & \cellcolor{lightgray}2 & \cellcolor{lightgray} & \cellcolor{lightgray} & \cellcolor{lightgray}0 & \cellcolor{lightgray}0 & \cellcolor{lightgray}0 & \cellcolor{lightgray}72 \\
\cellcolor{lightgray} & \cellcolor{lightgray}2 & \cellcolor{lightgray} & \cellcolor{lightgray}\checkmark & \cellcolor{lightgray}0 & \cellcolor{lightgray}0 & \cellcolor{lightgray}0 & \cellcolor{lightgray}40 \\
\cellcolor{lightgray} & \cellcolor{lightgray}1 & \cellcolor{lightgray} & \cellcolor{lightgray} & \cellcolor{lightgray}0 & \cellcolor{lightgray}0 & \cellcolor{lightgray}0 & \cellcolor{lightgray}72 \\
\cellcolor{lightgray} & \cellcolor{lightgray}1 & \cellcolor{lightgray} & \cellcolor{lightgray}\checkmark & \cellcolor{lightgray}0 & \cellcolor{lightgray}0 & \cellcolor{lightgray}0 & \cellcolor{lightgray}40 \\
\midrule
\cellcolor{white}Total & \cellcolor{white} & \cellcolor{white} & \cellcolor{white} & \cellcolor{white}241 & \cellcolor{white}148 & \cellcolor{white}43 & \cellcolor{white}224 \\
\bottomrule
\end{tabular}
\label{tab:data}
\end{table}

\vspace{-0.5cm}

\subsection{Comparative study}
We employed the nnU-Net framework~\cite{isensee_automated_2019, isensee_nnu-net_2024}, a fully automated, self-configuring DL method widely regarded as a state-of-the-art approach for biomedical image segmentation tasks. While nnU-Net provides automated optimization for key segmentation pipeline components, it allows for user modifications at critical stages. nnU-Net automatically optimizes the entire segmentation pipeline, including preprocessing, U-Net network architecture configuration, training procedure, inference, and post-processing, based solely on dataset characteristics. The framework's default process includes intensity normalization, resampling to optimal voxel spacing (resolution adjustment), cropping or padding to standardized patch sizes, and data augmentation (e.g., rotation, scaling, and elastic deformations) to enhance robustness.

Overall, the self-adaptive nature aims at nnU-Net's high segmentation accuracy and robust generalization across diverse clinical imaging applications. Moreover, previous work on CL segmentation used 3D U-Net variations~\cite{rosa_multiple_2020, la_rosa_multiple_2022} or nnU-Net~\cite{gordaliza_fluid_2025}. The default configuration of the framework employs a dynamically configured 3D U-Net architecture with an encoder-decoder structure and skip connections. Training typically uses a combined Dice and binary cross-entropy (BCE) loss function, stochastic gradient descent with momentum, and polynomial learning rate scheduling during 1000 epochs. The framework implements five-fold cross-validation, providing predictions as an ensemble of all folds. nnU-Net also offers automated post-processing that removes small isolated components below a threshold volume. 

In our systematic comparative experiments, we modified several key aspects of the default nnU-Net implementation to address the specific challenges of CL segmentation, like high class imbalance and the varied sizes of instances:

\paragraph{Optimization Strategy}
We substituted the default stochastic gradient descent with the Adam optimizer \cite{kingma_adam_2014} using a learning rate of 3e-4, which provided improved convergence behavior for our specific task.

\paragraph{Network Architecture} 
We systematically evaluated multiple architectural variants to address the unique challenges of CL segmentation. Beyond the baseline vanilla U-Net configuration provided by nnU-Net, we tested the recently introduced residual encoder variant that enhances gradient flow through deeper network layers. This residual encoder architecture is available in three configurations (designated "M", "L", and "XL") with progressively larger patch and batch sizes, allowing us to examine the trade-off between computational requirements and the ability to capture broader contextual information relevant to CL identification \cite{isensee_nnu-net_2024}. 
Furthermore, we explored the application of the U-Mamba model layers \cite{ma_u-mamba_2024} to the segmentation architecture. This approach was motivated by U-Mamba's theoretical advantages in modeling long-range spatial dependencies, which we hypothesized would better capture the relationship between lesion appearance and surrounding cortical anatomy. We tested two Mamba integration strategies: a conservative approach with Mamba layers only at the bottleneck, and a more extensive implementation with Mamba layers throughout the encoder pathway. 

\paragraph{Loss Function}
Beyond the default combination of Dice and cross-entropy losses, we evaluated its blob loss variant~\cite{kofler_blob_2023} to better capture the morphological characteristics of small CLs.

\paragraph{Resolution Standardization}
We modified the standard nnU-Net preprocessing pipeline by first upsampling all images to match the highest resolution dataset from site C (0.5mm³) using \textit{SimpleITK} with \textit{b-spline} interpolation. This preprocessing step ensured that when nnU-Net performed its automatic resolution optimization, it would determine 0.5mm³ as the optimal target resolution. This approach intends to preserve the fine structural details captured in the high-resolution site C dataset throughout the entire pipeline, potentially enhancing the detection of subtle cortical boundaries and lesions that might otherwise be obscured when nnU-Net's standard procedure would have selected a lower common-denominator resolution based on the heterogeneous input data.

\paragraph{Postprocessing} We found that the default postprocessing, removing small instances, is suboptimal for segmenting CL, as the small size of many CLs (e.g., a few voxels) meant that true positive lesions were being inappropriately removed.

We systematically evaluated framework modifications, including architectures, losses, and preprocessing, to quantify the improvement for CL detection on a dataset with diverse acquisition protocols and field strengths.

\subsection{Evaluation}

\subsubsection{Metrics}

We use a standard~\cite{la_rosa_cortical_2022, gordaliza_fluid_2025} evaluation pipeline, including the assessment of overall segmentation employing Dice similarity score (DSC), normalized DSC (nDSC)~\cite{raina_novel_2023}, and lesion detection quality (F1-score, Precision, and Recall).

Let TP, FP, FN be the number of true positive, false positive, and false negative voxel predictions. Then, the segmentation quality measures are defined as follows:

\[
DSC = \frac{2TP}{2TP + FP + FN},
\]

\[
nDSC = \frac{2TP}{2TP + \kappa \cdot FP + FN}, 
\quad \kappa = h(r^{-1} - 1),
\]

where $h$ is the ratio between the positive and the negative classes in the ground truth scan segmentation, 
$r \in (0, 1)$ is the reference value set to the mean fraction of the positive class. The nDSC accounts for the difference in the positive class load across different scans, addressing the class-load bias in the DSC.

Let TPL, FPL, and FNL be the number of true-positive, false-positive, and false-negative lesions. The lesions were classified on TPL, FPL, and FNL through the connected components analysis with 26 connectivity. A predicted connected component is a TPL if it has a one-voxel overlap with the ground truth; FPL if no overlap. A ground truth connected component is an FNL if it has no overlap with the prediction. Then, the detection quality measures are defined as follows:
\[
F_1\text{-score} = \frac{2TPL}{2TPL + FPL + FNL},
\]

\[
\text{Precision} = \frac{TPL}{TPL + FPL},
\quad
\text{Recall} = \frac{TPL}{TPL + FNL}.
\]

To assess the statistical significance of the differences in the medians of quality metrics for different models, the paired Wilcoxon tests on ranks were used. The p-values from multiple Wilcoxon tests were corrected using a Benjamini-Hochberg procedure for controlling the false discovery rate (FDR) with a 0.05 error rate. 

\subsubsection{Explainability analysis}
\textbf{Characterization of lesion-detection errors.} Sites A, B, and C manual annotations included different lesion phenotyping (leukocortical, subpial, WML, etc.). We used this information to better characterize the model errors. For this, we analyzed the overlaps between TPL, FPL, and FNL masks with labeled cortical and WM masks to obtain lesion classification. Additionally, we examined the relationship between lesion volume and errors.

\textbf{Bottleneck features analysis.} To gain insight into the internal representations learned by the network and to assess whether the model captures clinically or technically relevant variability, we performed an analysis of internal model representations on the bottleneck features extracted from the trained segmentation model. These features, located at the network's deepest layer, encode compressed information before upsampling and prediction. This analysis enables the exploration of the influence of data characteristics such as acquisition site, imaging modality, time point, or lesion burden on the learned feature space. Understanding these relationships can help uncover potential biases in the model and provide interpretability for its behavior across diverse data sources. Such information might be crucial given the diversity of the training data used in this study.

Specifically, we focused on the best-performing model and extracted its bottleneck features from the training dataset during inference. Since the model is an ensemble, the extracted nnU-Net bottleneck features were averaged across folds. Then, the bottleneck tensors were five-dimensional, shaped as (batch, channel, height, width, depth), where the batch dimension corresponds to augmented versions of the same input scan generated through test-time augmentation. To obtain the required (samples, features) format for dimensionality reduction, we averaged across the batch dimension and flattened the remaining spatial and channel dimensions, yielding a compact feature vector for each input scan.

We then applied dimensionality reduction to project these feature vectors into a three-dimensional space for visualization and further analysis. Two complementary techniques were employed: principal component analysis (PCA) and uniform manifold approximation and projection (UMAP). PCA is a linear method that identifies orthogonal directions of maximal variance in the data, allowing for efficient compression while preserving global structure~\cite{bishop_pattern_2006}. In contrast, UMAP is a nonlinear technique that constructs a low-dimensional embedding by preserving local neighborhood relationships, often capturing subtle nonlinear structures in the data~\cite{mcinnes_umap_2020}. The use of both methods allowed us to contrast global variance-driven patterns with local manifold structures. The dimensionality reduction models were fitted exclusively on the training data and subsequently used to project the test data into the same space. This was done primarily to understand the specifics of the trained features and analyze their relationship to the testing data.

The resulting embeddings were colored and grouped according to categorical variables (such as medical center, MRI modality, and field strength). These visualizations provided a qualitative assessment of how strongly the model's internal representation space reflects these underlying characteristics, indicating the model's capacity to differentiate or generalize across different domains.

\section{Results}

\subsection{Comparative study}

\subsubsection{Experiment 1: architectures comparison}

\begin{figure}[H]
    \centering
    \caption{Radial plots comparing different architectures using quality metric means and 90\% CIs. P-values from the paired Wilcoxon statistical tests with FDR correction, comparing model X with the baseline, i.e., Default resampling (BCE + Dice Loss): * p < 0.05, ** p < 0.01, *** p < 0.001 compared to the Vanilla baseline model.}
    \subfloat[Test-in]{
        \includegraphics[width=0.45\textwidth]{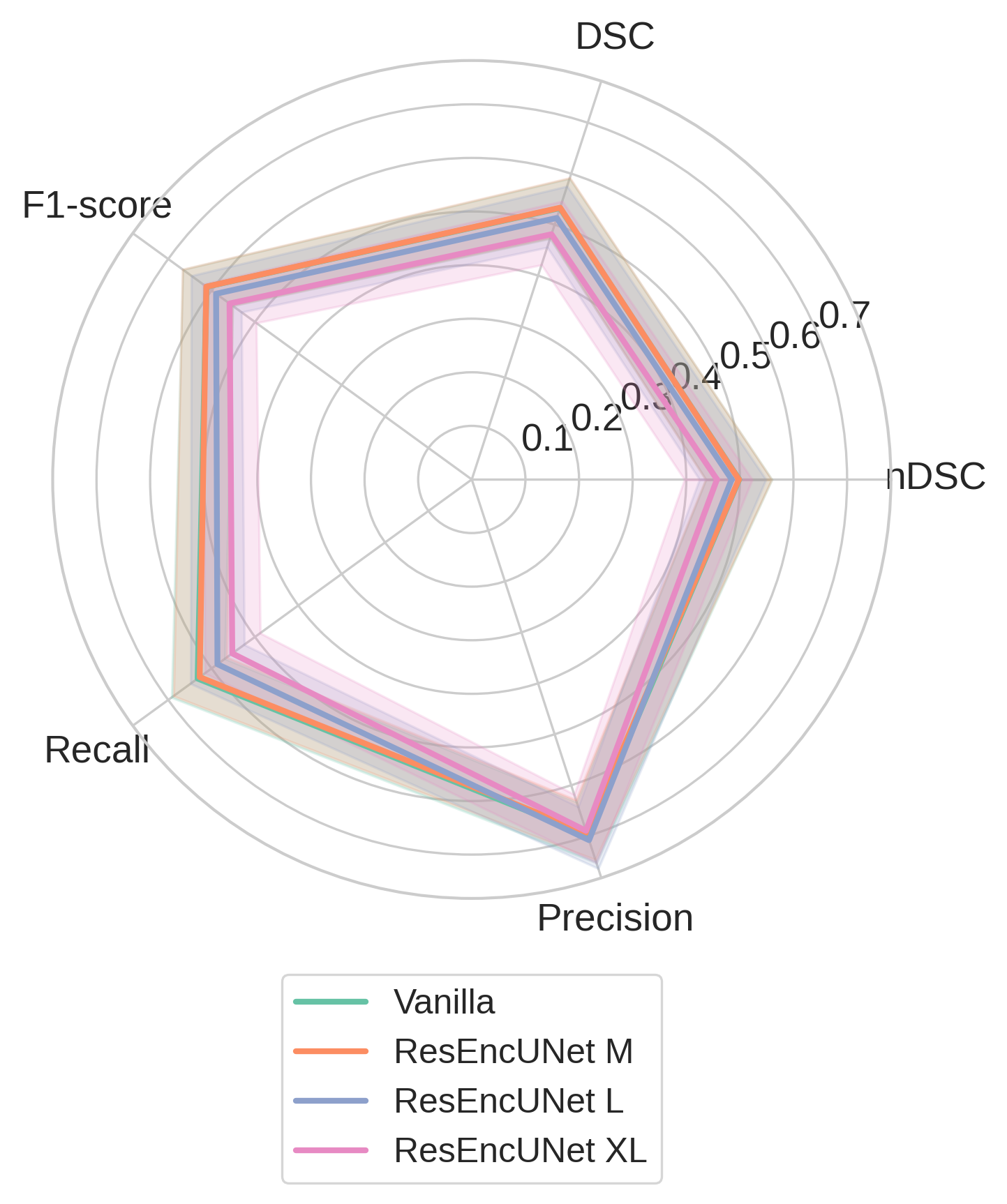} 
    }
    \hfill
    \subfloat[Test-out]{
        \includegraphics[width=0.45\textwidth]{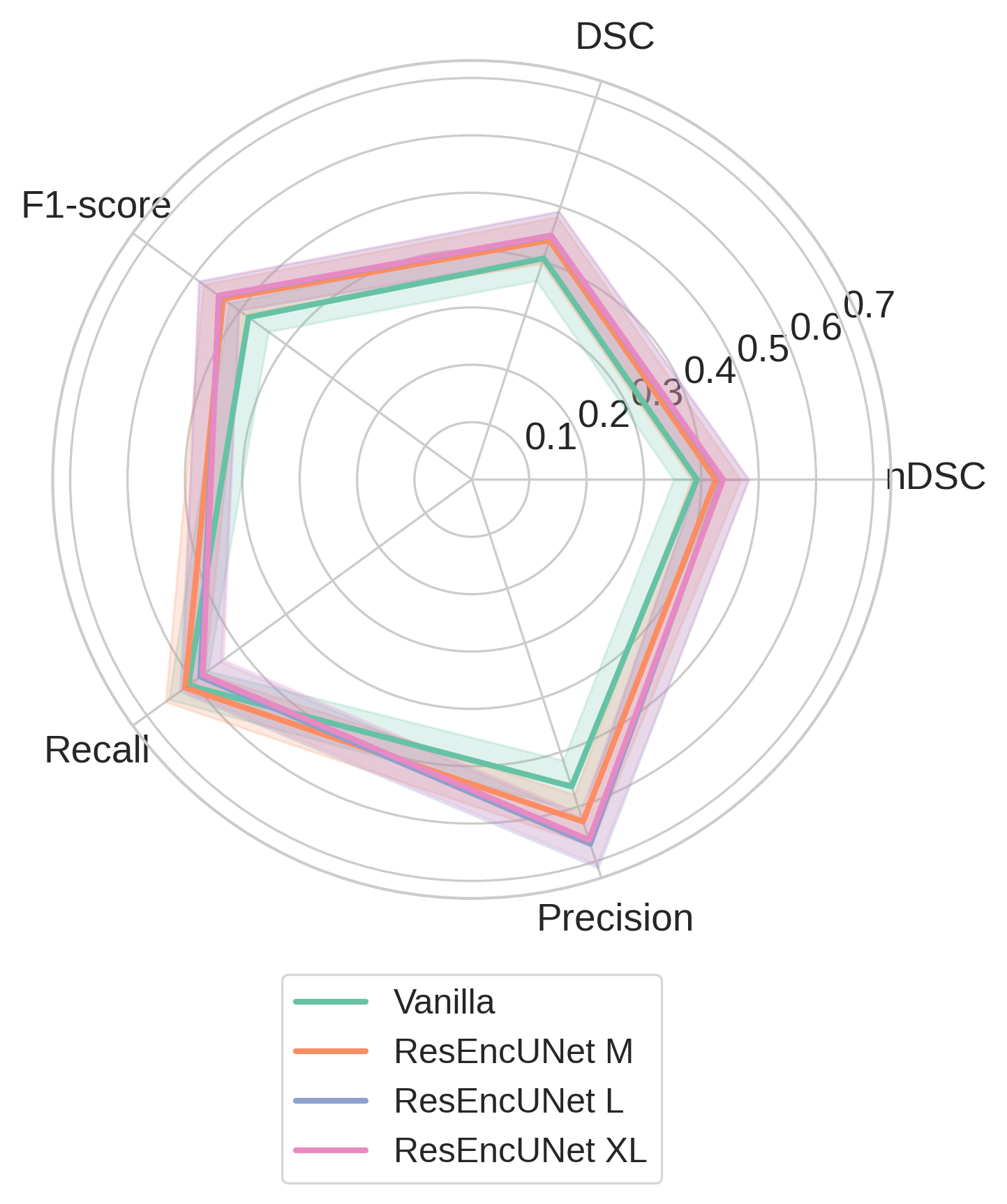}
    }
    \label{fig:e1rp}
\end{figure}

The model performance for different architectures is shown in Figure \ref{fig:e1rp}, and the values of the metrics with standard errors are shown in Table \ref{tab:e1ti}. 

U-Mamba architectures could not finally be included in the comparative analysis as we encountered significant convergence challenges during training that prevented reliable model fitting despite multiple parameter configurations and optimization attempts. These convergence issues persisted across both the bottleneck-only and full-encoder implementations of the U-Mamba layers, aligning with known training instabilities in state-space models \cite{isensee_nnu-net_2024}.

Among the successfully trained architectures, the performance differences were minimal. As shown in Tables \ref{tab:e1ti}, the Vanilla architecture achieved the highest normalized DSC ($0.499 \pm 0.038$) and recall ($0.632 \pm 0.038$) in the \textit{in-domain (Test-in)} dataset. At the same time, ResEncUNet variants had insignificantly fewer false-positive lesions on the \textit{out-of-domain (Test-out)} dataset. Since the differences between the models were not significant, the choice of more computationally expensive larger architectures was not justified. Thus, we used the Vanilla architecture for further experiments.

\vspace{-1cm}
\begin{table}[H]
\caption{Performance comparison of model architectures across i) Test-in and ii) Test-out domains. Values represent mean performance with standard error as subscripts. Bold values indicate the highest performance for each metric.}
\centering
\scriptsize
\label{tab:e1ti}
\subfloat[Test-in]{
    \begin{tabular}{p{2.5cm}|*{5}{p{1.2cm}}} 
    \toprule
    \rowcolor{headercolor}
    \textbf{Model} & \textbf{nDSC} & \textbf{DSC} & \textbf{F1-score} & \textbf{Recall} & \textbf{Precision} \\
    \midrule
    \cellcolor{lightgray}\textbf{{Vanilla}} & \cellcolor{lightgray}\textbf{0.499}$_{0.038}$ & \cellcolor{lightgray}0.532$_{0.036}$ & \cellcolor{lightgray}0.612$_{0.035}$ & \cellcolor{lightgray}\textbf{0.632}$_{0.038}$ & \cellcolor{lightgray}0.698$_{0.036}$ \\
    \cellcolor{white}\textbf{{ResEncUNet M}} & \cellcolor{white}0.498$_{0.038}$ & \cellcolor{white}\textbf{0.533}$_{0.036}$ & \cellcolor{white}\textbf{0.612}$_{0.036}$ & \cellcolor{white}0.627$_{0.038}$ & \cellcolor{white}0.693$_{0.036}$ \\
    \cellcolor{lightgray}\textbf{{ResEncUNet L}} & \cellcolor{lightgray}0.485$_{0.038}$ & \cellcolor{lightgray}0.513$_{0.037}$ & \cellcolor{lightgray}0.590$_{0.035}$ & \cellcolor{lightgray}0.586$_{0.038}$ & \cellcolor{lightgray}\textbf{0.708}$_{0.037}$ \\
    \cellcolor{white}\textbf{{ResEncUNet XL}} & \cellcolor{white}0.457$_{0.039}$ & \cellcolor{white}0.481$_{0.038}$ & \cellcolor{white}0.559$_{0.038}$ & \cellcolor{white}0.552$_{0.039}$ & \cellcolor{white}0.689$_{0.041}$ \\
    \bottomrule
    \end{tabular}
}
\qquad
\subfloat[Test-out]{
    \begin{tabular}{p{2.5cm}|*{5}{p{1.2cm}}} 
    \toprule
    \rowcolor{headercolor}
    \textbf{Model} & \textbf{nDSC} & \textbf{DSC} & \textbf{F1-score} & \textbf{Recall} & \textbf{Precision} \\
    \midrule
    \cellcolor{lightgray}\textbf{{Vanilla}} & \cellcolor{lightgray}0.392$_{0.025}$ & \cellcolor{lightgray}0.405$_{0.025}$ & \cellcolor{lightgray}0.481$_{0.026}$ & \cellcolor{lightgray}0.611$_{0.026}$ & \cellcolor{lightgray}0.563$_{0.029}$ \\
    \cellcolor{white}\textbf{{ResEncUNet M}} & \cellcolor{white}0.425$_{0.026}$ & \cellcolor{white}0.439$_{0.026}$ & \cellcolor{white}0.536$_{0.026}$ & \cellcolor{white}\textbf{0.618}$_{0.025}$ & \cellcolor{white}0.627$_{0.028}$ \\
    \cellcolor{lightgray}\textbf{{ResEncUNet L}} & \cellcolor{lightgray}0.436$_{0.027}$ & \cellcolor{lightgray}0.447$_{0.026}$ & \cellcolor{lightgray}0.544$_{0.026}$ & \cellcolor{lightgray}0.585$_{0.026}$ & \cellcolor{lightgray}\textbf{0.668}$_{0.029}$ \\
    \cellcolor{white}\textbf{{ResEncUNet XL}} & \cellcolor{white}\textbf{0.438}$_{0.027}$ & \cellcolor{white}\textbf{0.447}$_{0.027}$ & \cellcolor{white}\textbf{0.545}$_{0.026}$ & \cellcolor{white}0.579$_{0.027}$ & \cellcolor{white}0.661$_{0.028}$ \\
    \bottomrule
    \end{tabular}
}
\end{table}

\subsubsection{Experiment 2: Blob loss effect}
The Vanilla model performance for different loss functions is illustrated in Figure \ref{fig:e23rp}, and the values of the metrics with standard errors are shown in Tables \ref{tab:e23ti}. 

Blob BCE + Dice Loss consistently outperformed standard BCE + Dice Loss across most metrics. For test-in domain, improvements included nDSC (0.505 vs 0.499), DSC (0.539 vs 0.532), F1-score (0.641 vs 0.612), Precision (0.715 vs 0.698). Similar improvements were observed out-of-domain, with the most notable gain in Precision (0.592 vs 0.563, p < 0.05). Statistical significance was achieved for out-of-domain Precision.

\subsubsection{Experiment 3: effect of image upsampling}
\begin{figure}[H]
    \centering
    \caption{Radial plots comparing different losses and resampling strategies using model performance metrics means and 90\% CIs. P-values from the paired Wilcoxon statistical tests with FDR correction, comparing model X with the baseline, i.e., Vanilla: * p < 0.05, ** p < 0.01, *** p < 0.001 compared to baseline model.}
    \subfloat[Test-in]{
        \includegraphics[width=0.45\textwidth]{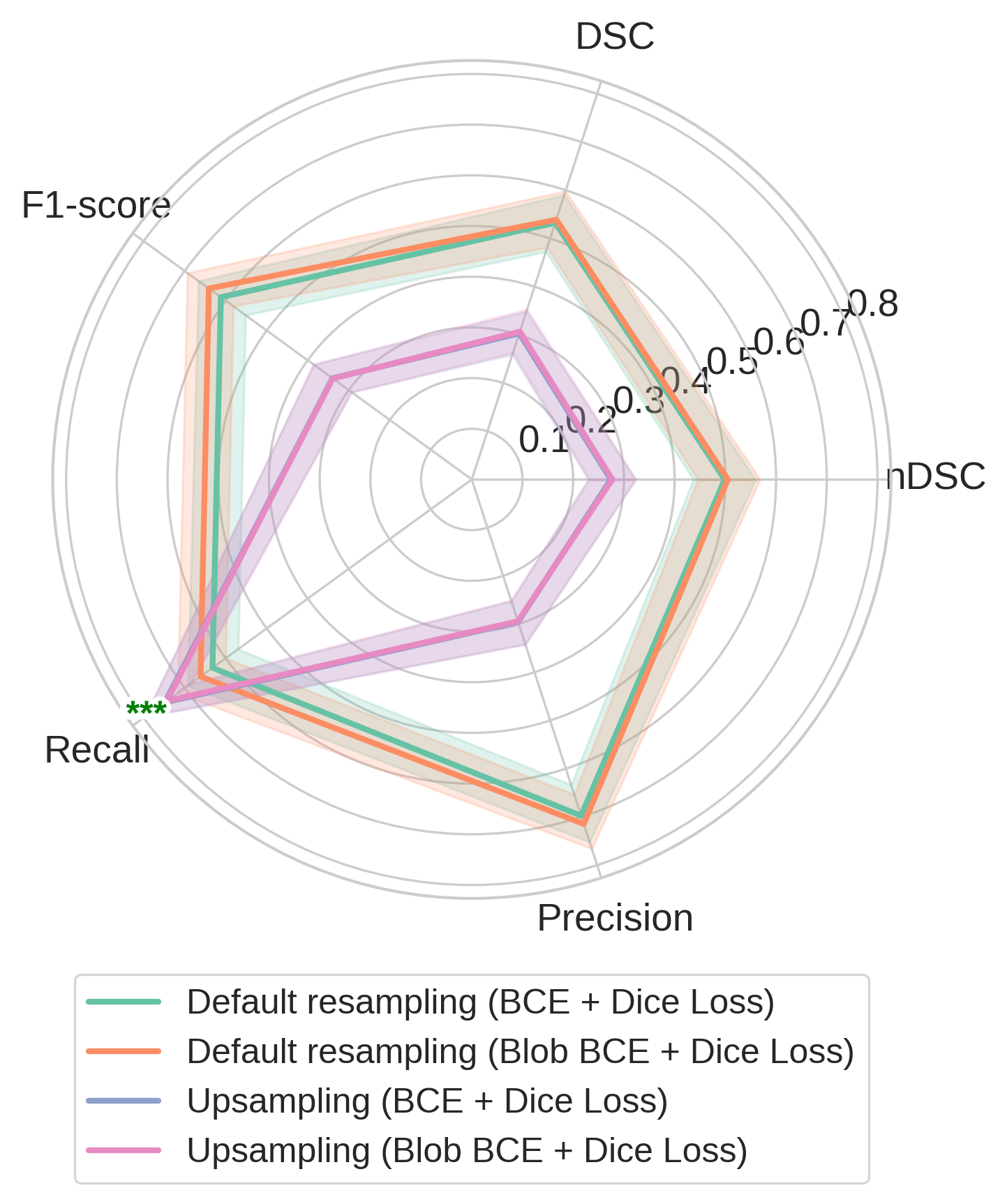} 
    }
    \hfill
    \subfloat[Test-out]{
        \includegraphics[width=0.45\textwidth]{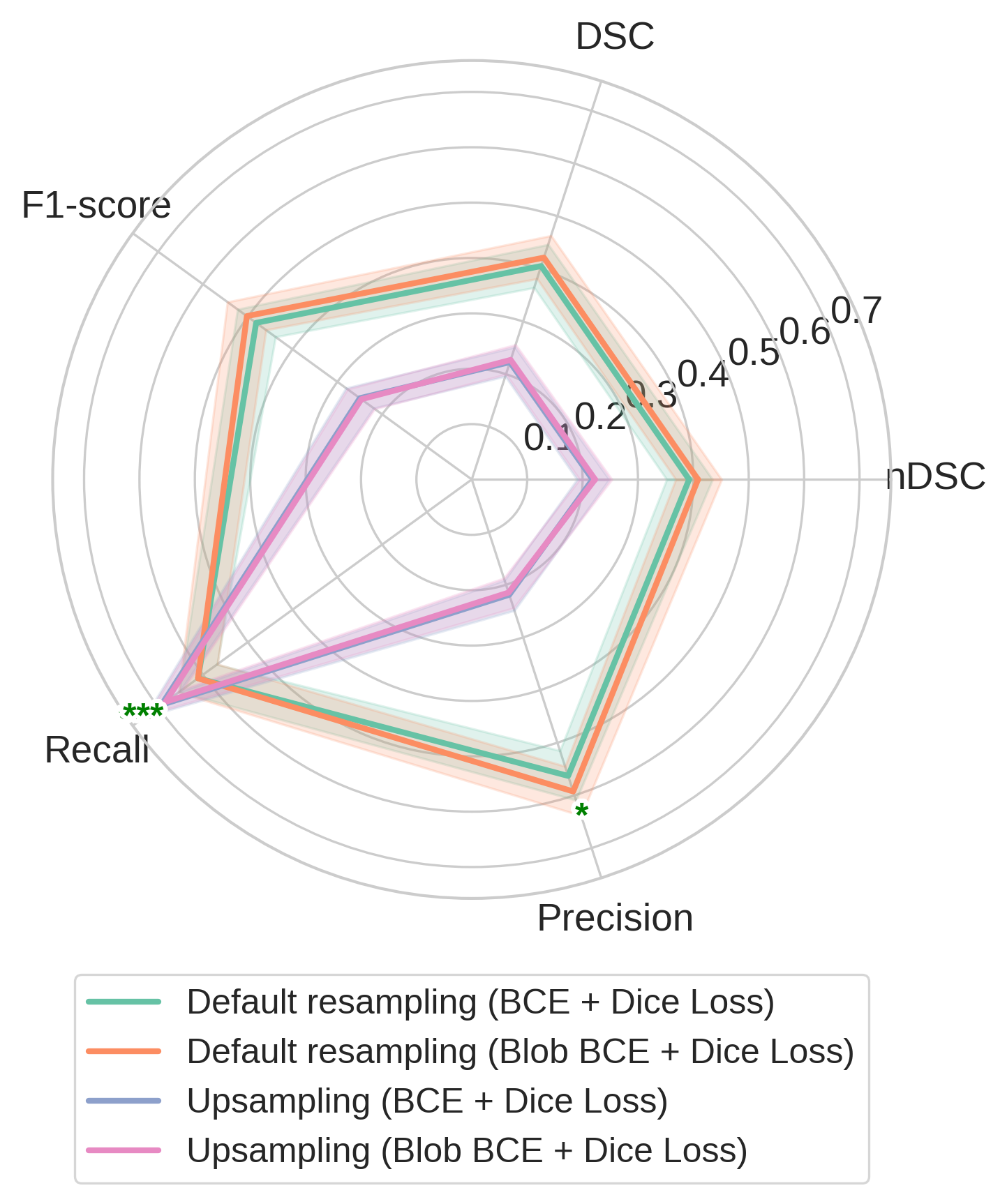} 
    }
    \label{fig:e23rp}
\end{figure}

The model performance for different image resampling strategies is illustrated in Figure \ref{fig:e23rp}, and the values of the metrics with standard errors are shown in Tables \ref{tab:e23ti}.

Upsampling significantly improved lesion detection, increasing Recall from 0.632 to 0.748 (p < 0.001) for test-in domain. However, this came at the cost of substantially increased false positives, with Precision dropping from 0.698 to 0.296. The poor precision-recall trade-off resulted in overall performance degradation, with F1-scores decreasing by approximately 45\% (from 0.641 to 0.339). 

\vspace{-1cm}
\begin{table}[H]
\caption{Performance comparison of loss functions with and without upsampling across i) Test-in and ii) Test-out domains. Values represent mean performance with standard error as subscripts. Bold values indicate the highest performance for each metric.}
\centering
\scriptsize
\label{tab:e23ti}
\subfloat[Test-in]{
    \begin{tabular}{p{2.5cm}|c|*{5}{p{1.2cm}}}
    \toprule
    \rowcolor{headercolor}
    \textbf{Loss function} & \textbf{Up-sampling} & \textbf{nDSC} & \textbf{DSC} & \textbf{F1-score} & \textbf{Recall} & \textbf{Precision} \\
    \midrule
    \multirow{2}{*}{\centering BCE + Dice} & \cellcolor{lightgray} & \cellcolor{lightgray}$0.499_{0.038}$ & \cellcolor{lightgray}$0.532_{0.036}$ & \cellcolor{lightgray}$0.612_{0.035}$ & \cellcolor{lightgray}$0.632_{0.038}$ & \cellcolor{lightgray}$0.698_{0.036}$ \\
    & \cellcolor{white}\checkmark & \cellcolor{white}$0.274_{0.029}$ & \cellcolor{white}$0.303_{0.028}$ & \cellcolor{white}$0.340_{0.028}$ & \cellcolor{white}\textbf{$0.748_{0.033}$} & \cellcolor{white}$0.296_{0.028}$ \\
    \midrule
    \multirow{2}{*}{\centering Blob BCE + Dice} & \cellcolor{lightgray} & \cellcolor{lightgray}\textbf{$0.505_{0.038}$} & \cellcolor{lightgray}\textbf{$0.539_{0.036}$} & \cellcolor{lightgray}\textbf{$0.641_{0.034}$} & \cellcolor{lightgray}$0.661_{0.036}$ & \cellcolor{lightgray}\textbf{$0.715_{0.034}$} \\
    & \cellcolor{white}\checkmark & \cellcolor{white}$0.277_{0.028}$ & \cellcolor{white}$0.307_{0.028}$ & \cellcolor{white}$0.339_{0.029}$ & \cellcolor{white}$0.743_{0.034}$ & \cellcolor{white}$0.294_{0.028}$ \\
    \bottomrule
    \end{tabular}
}
\qquad
\subfloat[Test-out]{
    \begin{tabular}{p{2.5cm}|c|*{5}{p{1.2cm}}}
    \toprule
    \rowcolor{headercolor}
    \textbf{Loss function} & \textbf{Up-sampling} & \textbf{nDSC} & \textbf{DSC} & \textbf{F1-score} & \textbf{Recall} & \textbf{Precision} \\
    \midrule
    \multirow{2}{*}{\centering BCE + Dice} & \cellcolor{lightgray} & \cellcolor{lightgray}$0.392_{0.025}$ & \cellcolor{lightgray}$0.405_{0.025}$ & \cellcolor{lightgray}$0.481_{0.026}$ & \cellcolor{lightgray}$0.611_{0.026}$ & \cellcolor{lightgray}$0.563_{0.029}$ \\
    & \cellcolor{white}\checkmark & \cellcolor{white}$0.218_{0.017}$ & \cellcolor{white}$0.223_{0.017}$ & \cellcolor{white}$0.250_{0.019}$ & \cellcolor{white}\textbf{$0.692_{0.023}$} & \cellcolor{white}$0.219_{0.017}$ \\
    \midrule
    \multirow{2}{*}{\centering Blob BCE + Dice} & \cellcolor{lightgray} & \cellcolor{lightgray}\textbf{$0.408_{0.026}$} & \cellcolor{lightgray}\textbf{$0.421_{0.025}$} & \cellcolor{lightgray}\textbf{$0.502_{0.026}$} & \cellcolor{lightgray}$0.611_{0.026}$ & \cellcolor{lightgray}\textbf{$0.592_{0.029}$} \\
    & \cellcolor{white}\checkmark & \cellcolor{white}$0.223_{0.018}$ & \cellcolor{white}$0.227_{0.017}$ & \cellcolor{white}$0.246_{0.019}$ & \cellcolor{white}$0.683_{0.024}$ & \cellcolor{white}$0.215_{0.017}$ \\
    \bottomrule
    \end{tabular}
}
\end{table}

\subsection{Best model analysis}
We selected the Vanilla nnU-net with Blob loss model for a more detailed performance evaluation and further explainability analysis.

\subsubsection{Details on the model performance}
The model performance in terms of lesion segmentation and detection across different sites, modalities, and diseases is shown in Table \ref{tab:bmp}. 
According to the Kruskal-Wallis H-test with FDR correction, the differences between sites and sites $\times$ modality are significant for all metrics, except for the precision: adjusted p-value=0.384 for precision and < 0.008 for other metrics. 
Among the MS patients, site A has the highest DSC and detection F1. 
Around a 30\% drop in DSC and F1 was observed for site B compared to site A. 
Within site B, the difference between MPRAGE and MP2RAGE is marginally small and is not significant according to the U-test (adjusted p-value = 0.877 for DSC and 0.933 for F1). 
An additional 50\% drop in DSC and F1 was observed for the site C 7T data; however, it only had 5 testing cases. FNL should largely explain this drop in performance, since the number is outstandingly high. 
OOD lesion detection performance was affected by a larger number of false negatives: \#FNL in sites A and B was around 3 lesions per scan, in site D, 6.5 lesions per scan. 
We compared the OOD site D with the in-domain sites (A, B, C) performance using the Mann-Whitney U-test with FDR correction of p-values, and detected a significant difference in the nDSC, DSC, and F1 with adjusted p-values of 0.015, 0.004, and 0.007, respectively. However, the precision, recall, and FPL and FNL counts are not significantly different.

\begin{table}[!hb]
\caption{Segmentation and lesion detection quality for the best model (Vanilla nnU-Net with Blob loss) per site and modality (1 - MPRAGE, 2 - MP2RAGE; 7T if ultra-high field, else 3T). Site D has a separation on MS and MS-mimics. Mean average and standard error were computed across subjects from Test-in (A, B, C) and Test-out (D).}
\centering
\scriptsize
\label{tab:bmp}
\subfloat[Segmentation quality]{
\begin{tabular}{>{\columncolor{white}}p{0.8cm} p{0.8cm} p{0.8cm} p{0.5cm} p{0.5cm} | p{1.3cm} p{1.3cm}}
\toprule
\rowcolor{headercolor}
    \textbf{Site} & \textbf{MS-mimic} & \textbf{Moda-lity} & \textbf{FS} & \textbf{n} & \textbf{nDSC} & \textbf{DSC} \\
\midrule
\rowcolor{lightgray}
\multirow{-1}{*}{\centering A} & & 2 & & 50 & 0.604$_{0.046}$ & 0.627$_{0.044}$ \\
\midrule
\multirow{-1}{*}{\centering B} 
& & 1 &  & 14 & 0.445$_{0.090}$ & 0.473$_{0.092}$ \\
& & 2 &  & 14 & 0.420$_{0.089}$ & 0.431$_{0.090}$ \\
\midrule
\rowcolor{lightgray}
\multirow{-1}{*}{\centering C} & & 2 &  & 1 & 0.215$_{0.000}$ & 0.404$_{0.000}$ \\
\rowcolor{lightgray}
& & 2 & \checkmark & 6 & 0.068$_{0.012}$ & 0.233$_{0.059}$ \\
\midrule
\multirow{-1}{*}{\centering D} 
& & 1 &  & 72 & 0.298$_{0.028}$ & 0.319$_{0.029}$ \\
& & 2 &  & 72 & 0.292$_{0.030}$ & 0.311$_{0.029}$ \\
& \checkmark & 1 &  & 40 & 0.600$_{0.078}$ & 0.600$_{0.078}$ \\
& \checkmark & 2 &  & 40 & 0.625$_{0.078}$ & 0.625$_{0.078}$ \\
\bottomrule
\end{tabular}
}
\qquad
\subfloat[Lesion detection quality]{
\begin{tabular}{>{\columncolor{white}}p{0.8cm} p{0.8cm} p{0.8cm} p{0.5cm} p{0.5cm} | *{3}{p{1.3cm}} p{1cm} p{1cm}} 
\toprule
\rowcolor{headercolor}
\textbf{Site} & \textbf{MS-mimic} & \textbf{Moda-lity} & \textbf{FS} & \textbf{n} & \textbf{F1-score} & \textbf{Recall} & \textbf{Precision} & \textbf{\#FPL} & \textbf{\#FNL} \\\midrule
\rowcolor{lightgray}
\multirow{-1}{*}{\centering A} & & 2 &  & 50 & 0.742$_{0.038}$ & 0.783$_{0.039}$ & 0.741$_{0.041}$ & 1.9$_{0.3}$ & 2.7$_{0.6}$ \\
\midrule
\multirow{-1}{*}{\centering B} 
& & 1 &  & 14 & 0.536$_{0.095}$ & 0.554$_{0.097}$ & 0.677$_{0.103}$ & 1.1$_{0.4}$ & 3.7$_{1.0}$ \\
& & 2 &  & 14 & 0.532$_{0.092}$ & 0.538$_{0.090}$ & 0.679$_{0.107}$ & 1.0$_{0.4}$ & 3.6$_{0.9}$ \\
\midrule
\rowcolor{lightgray}
\multirow{-1}{*}{\centering C} & & 2 &  & 1 & 0.593$_{0.000}$ & 0.471$_{0.000}$ & 0.800$_{0.000}$ & 2.0$_{0.0}$ & 9.0$_{0.0}$ \\
\rowcolor{lightgray}
& & 2 & \checkmark & 6 & 0.302$_{0.062}$ & 0.216$_{0.051}$ & 0.645$_{0.060}$ & 7.0$_{3.1}$ & 55.5$_{15.9}$ \\
\midrule
\multirow{-1}{*}{\centering D} 
& & 1 &  & 72 & 0.446$_{0.038}$ & 0.395$_{0.038}$ & 0.586$_{0.046}$ & 1.8$_{0.4}$ & 6.5$_{1.3}$ \\
& & 2 &  & 72 & 0.435$_{0.037}$ & 0.395$_{0.038}$ & 0.577$_{0.047}$ & 1.7$_{0.4}$ & 6.7$_{1.4}$ \\
& \checkmark & 1 &  & 40 & 0.600$_{0.078}$ & - & 0.600$_{0.078}$ & 1.5$_{0.7}$ & - \\
& \checkmark & 2 &  & 40 & 0.625$_{0.078}$ & - & 0.625$_{0.078}$ & 0.8$_{0.3}$ & - \\
\bottomrule
\end{tabular}
}
\end{table}

\subsubsection{Lesion-detection errors analysis}
\begin{figure}[h!]
    \centering
    \includegraphics[width=0.73\linewidth]{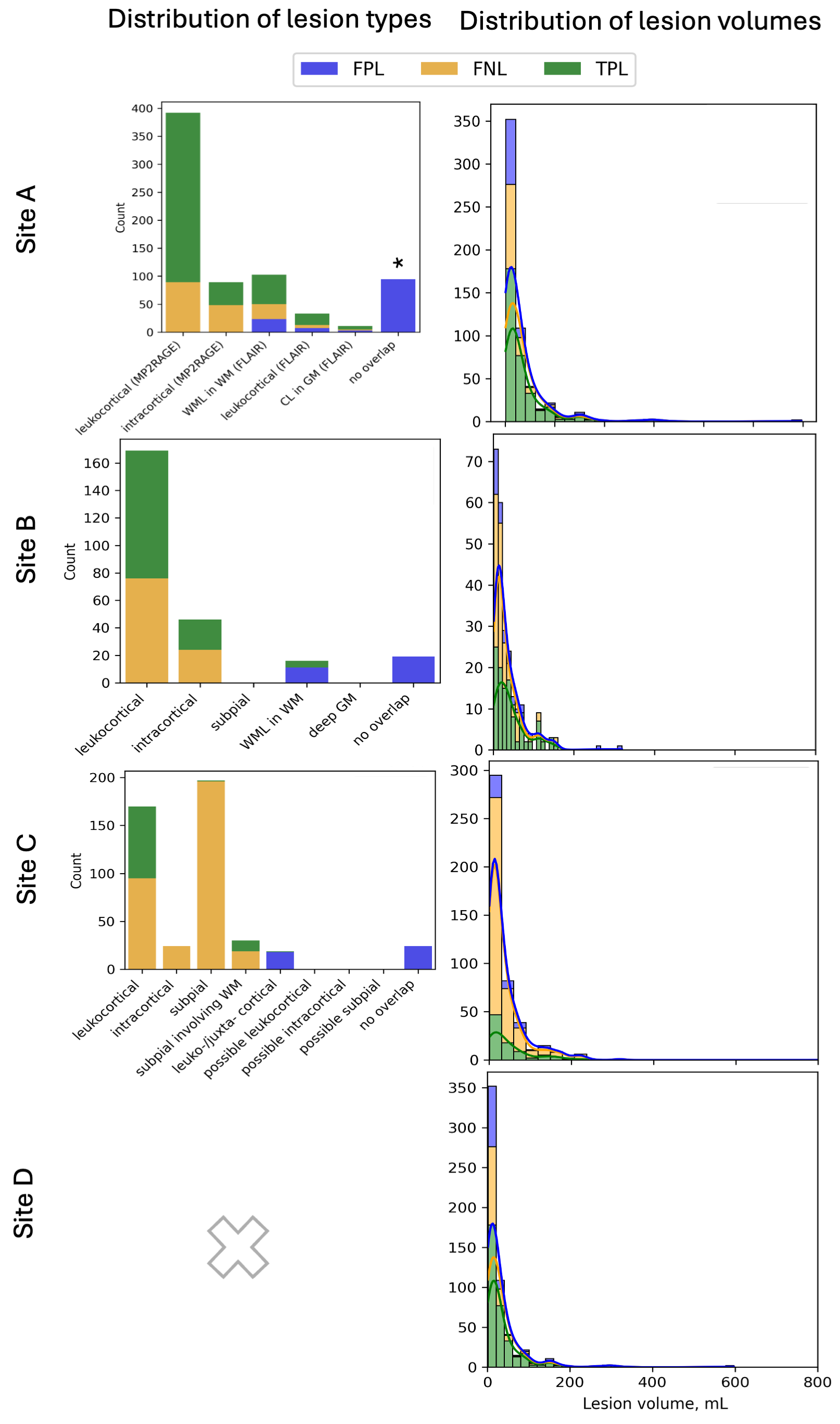}    \caption{Distributions of lesion types and volumes across FPL, FNL, and TPL categories from sites A-D (rows).  $^{\star}$For site A WML segmentation was only available for timepoint 1, limiting FPL typization.}
    \label{fig:les_dist}
\end{figure}

The distribution of the lesion types and sizes across the TPL, FPL, and FNL groups is shown in Figure \ref{fig:les_dist}.

For both site A and B, we observed that the leukocortical lesions are more commonly undetected, compared to other types of lesions (65\% of all FNLs from site A and 75\% from site B). Also, we see that some FPLs had intersections with WML masks or leuko-/intra-cortical lesions annotated on FLAIR. Particularly, for site A timepoint 1, where the FLAIR annotations were available, 27 of 83 FPLs (33\%) fell into the category of WML in WM, 6 FPLs (7\%) were leukocortical under FLAIR, and 2 FPLs (2\%) into CL in GM under FLAIR (i.e., intracortical). For site C, out of 333 FNLs, 196 were subpial lesions (59 \% of all FNLs) and 95 were leukocortical lesions (29\%). Out of 42 FPLs, 18 fall into the category of uncertain leuko-/juxta-cortical lesions. For all the sites, the majority of TPL are leukocortical lesions.

The distribution of lesion volumes communicates that the majority of missed lesions are small. Across sites A-C, the median FNL volume is 15mL, which is 6-18 mL less than the median TPL volume.

\begin{figure}[h!]
    \centering
    \includegraphics[width=0.73\linewidth]{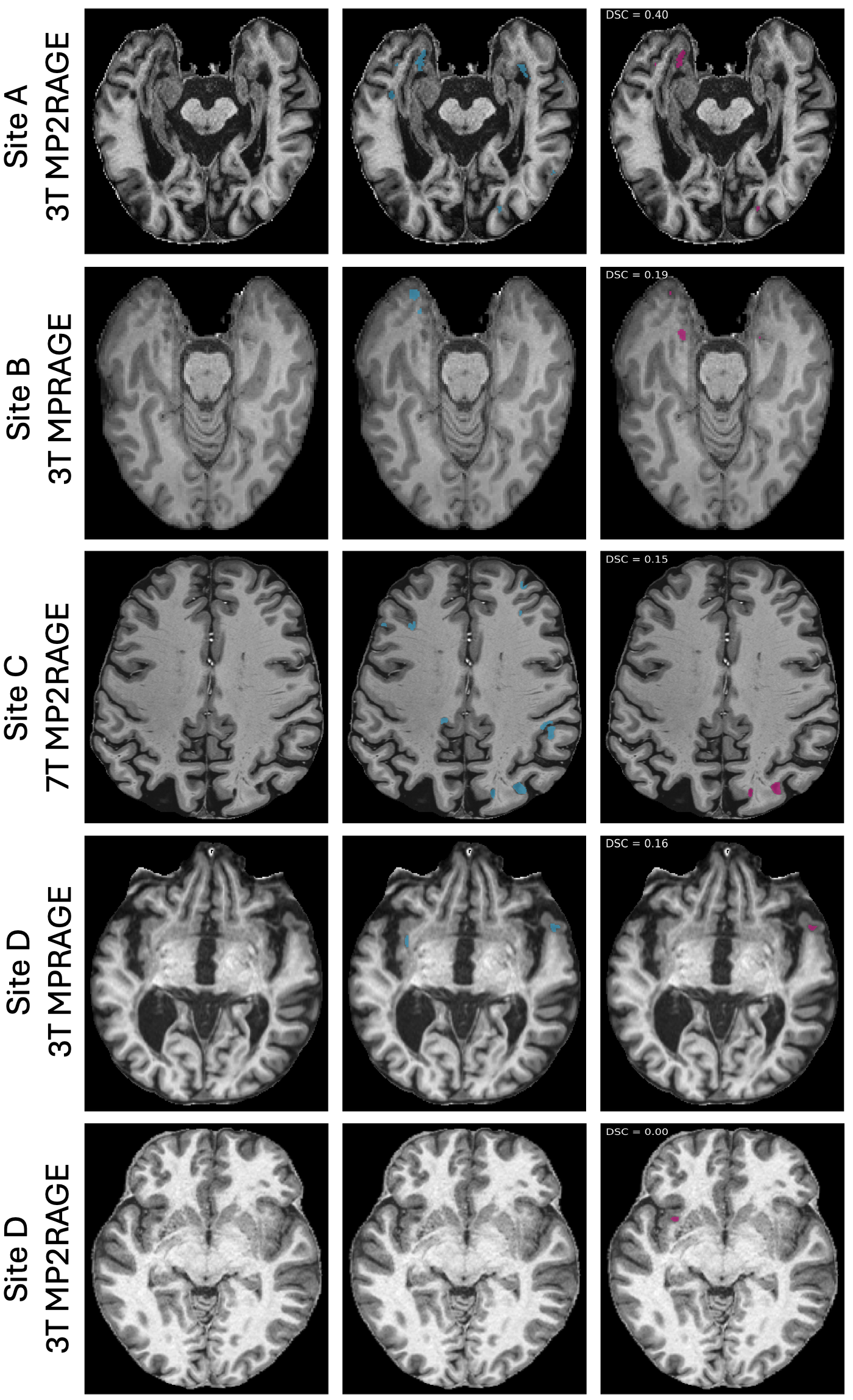}
    \caption{Visualization of the manual and automatic segmentation per site for scans with low segmentation quality (DSC < Q2) and median lesion count per site. Images from left to right show MRI axial slices with i) no segmentation, ii) ground truth mask overlay (blue), iii) predicted mask overlay (pink). }
    \label{fig:low_perf}
\end{figure}

\begin{figure}[h!]
    \centering
    \includegraphics[width=0.8\linewidth]{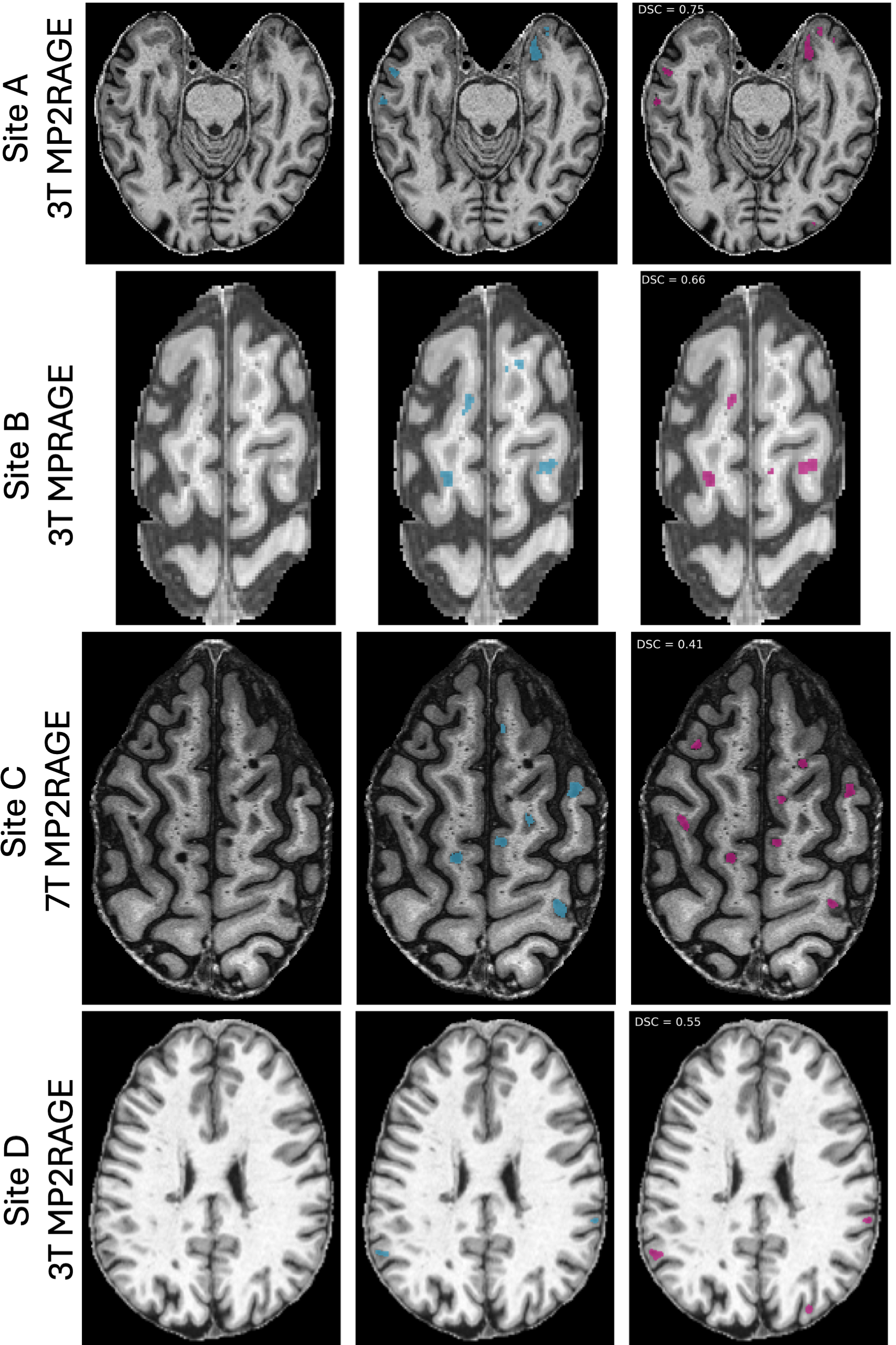}
    \caption{Visualization of the manual and automatic segmentation per site for scans with high segmentation quality (DSC > Q3) and median lesion count per site. Images from left to right show MRI axial slices with i) no segmentation, ii) ground truth mask overlay (blue), iii) predicted mask overlay (pink).}
    \label{fig:high_perf}
\end{figure}

For a qualitative assessment, we visualized subjects with low (DSC < second quartile - Q2) versus high (DSC > third quartile - Q3) segmentation quality for different sites in Figure \ref{fig:low_perf} and \ref{fig:high_perf}.

\subsubsection{Bottleneck features analysis}
\begin{figure}[h!]
\caption{Bottleneck features visualization using PCA (left) and UMAP (right) dimensionality reduction.}
    \centering
    \includegraphics[width=0.9\linewidth]{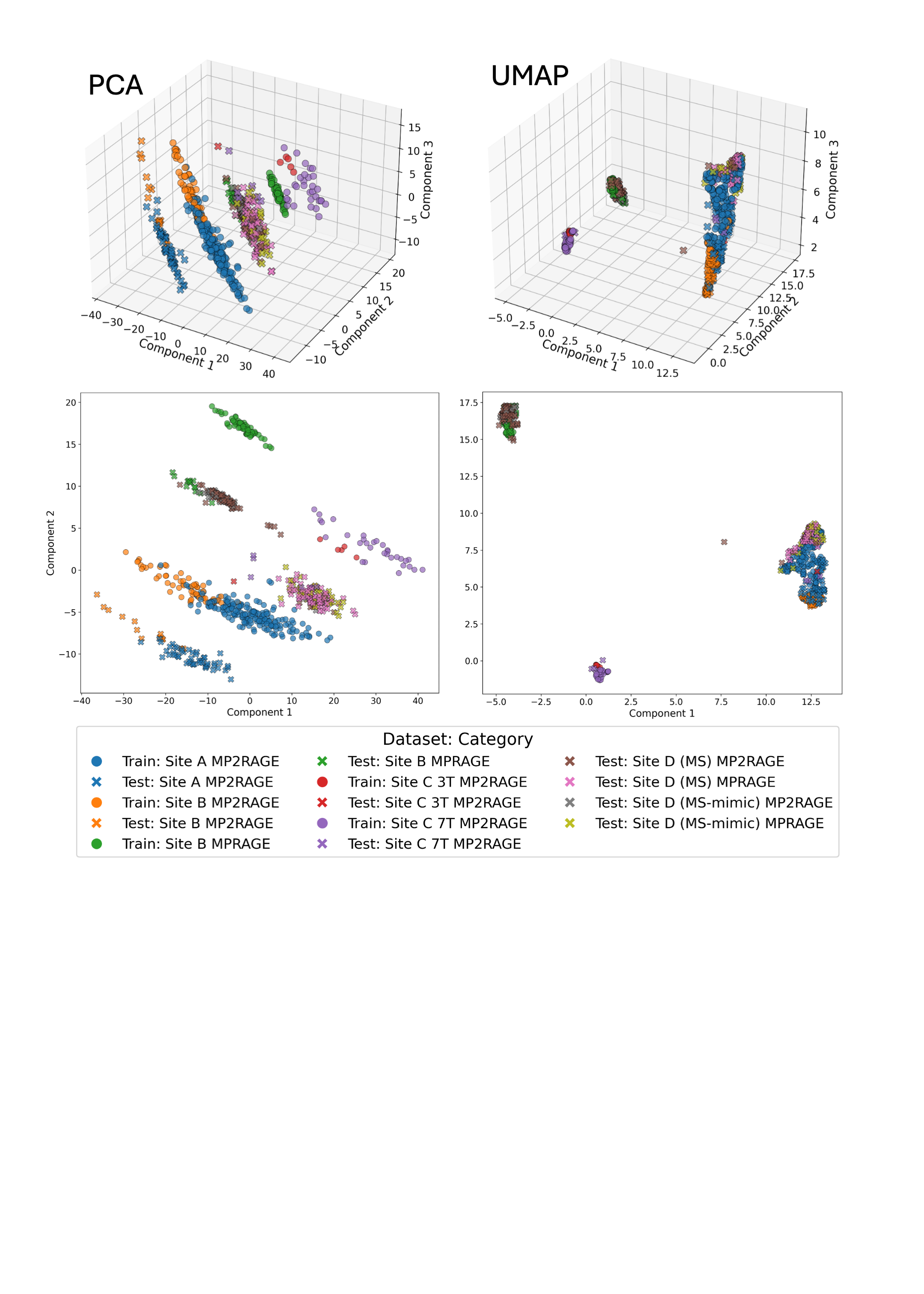}
    \label{fig:latent}
\end{figure}

The results of the nnU-Net bottleneck feature analysis are shown in Figure \ref{fig:latent} for PCA and UMAP methods. For PCA projection, the clusters formed partially correspond to the site $\times$ modality. The biggest cluster includes site A and site B MP2RAGE. The testing data from these sites is slightly shifted from the training data. For both PCA and UMAP, MPRAGE scans from site B and MP2RAGE scans from site C are grouped, which is due to the intensity distribution, given the relatively large distance between the MP2RAGE scans from site B of the same patients. For both PCA and UMAP, the data from site C form a distinct cluster, indicating its separation from the rest of the data. Most of the OOD test data, projected into the space of training bottleneck features, is located close to the largest cluster (sites A and B, MP2RAGE). For PCA, OOD test data has a central location for the first three components. Additional results quantifying the relationship between different continuous features, like volumetric errors, DSC, and uncertainty, with the bottleneck features are in the \ref{ap:latent}.

\section{Discussion}

\subsection{Model comparison}
Our experiments demonstrated that simpler neural network architectures, specifically the vanilla nnU-Net, yielded comparable or superior CL segmentation performance compared to larger architectures, which suffered from convergence difficulties. This observation aligns with prior studies advocating for lightweight models in medical image segmentation tasks. For instance, LV-UNet and Mini-Net have shown that smaller architectures can achieve competitive performance while being more efficient and easier to train \cite{jiang_lv-unet_2024, javed_advancing_2024}.

The customized Blob BCE + Dice Loss modestly improved precision, particularly for OOD data, reinforcing the importance of lesion-specific loss functions in optimizing segmentation tasks involving small lesions. A similar outcome was previously demonstrated for the WML segmentation on FLAIR MRI~\cite{kofler_blob_2023}. However, image resampling significantly improved lesion recall yet markedly increased false-positive detections, drastically reducing overall segmentation precision. B-spline MRI upsampling might lead to an increased number of artifacts, leading to higher false-positive rates. Given a promise of high recall, one could explore cascade architectures or two-step approaches to refine such segmentations with high false positive rates.

These findings collectively suggest that effective CL segmentation relies more on appropriate training strategies than architectural sophistication. The limited benefit of model complexity, combined with modest improvements achievable through specialized loss functions, has important implications for clinical deployment: simpler models offer substantial practical advantages, including reduced computational requirements, lower operational costs, and decreased energy consumption—critical factors for sustainable healthcare AI implementation \cite{topol_high-performance_2019, topol_as_2023}.


\subsection{Performance differences across sites}
Substantial segmentation performance differences were observed across imaging sites and MRI modalities. Segmentation accuracy was highest for MP2RAGE at 3T (sites A and B; DSC 0.47-0.63, detection F1 0.53-0.73), with slightly reduced performance for MPRAGE sequences at these same sites. Conversely, ultra-high-field (7T) data from site C displayed notably poorer performance down to 0.23 DSC and 0.29 detection F1. This drop is primarily attributable to the high prevalence of subpial lesions in site C, which are better detectable at ultra-high fields~\cite{maranzano_comparison_2019}. 
For sites A-C, leukocortical lesions were frequently missed (false negatives) by the automated segmentation approach, while numerous false positives corresponded to previously annotated WMLs. This difficulty reflects inherent ambiguity in distinguishing leukocortical from juxtacortical lesions, i.e., CL and WML, a well-known challenge in manual segmentation.

\subsection{Explainability analysis insights}
The apparent clustering observed in PCA projection—where testing data from sites A and B showed shifts from training distributions—may reflect the inherent nature of learned representations or true overfitting. PCA captures only linear relationships in the high-dimensional bottleneck features, and it is expected that certain feature channels will capture scanner-specific or site-specific characteristics (acquisition noise patterns, intensity distributions, contrast variations) alongside task-relevant CL representations. This scanner-specific information naturally creates linearly separable clusters in PCA space, even when the overall learned representation remains robust for the segmentation task. Therefore, PCA alone may not be sufficient for assessing model generalization, as it cannot distinguish between harmful overfitting and expected technical variation capture. The contrasting patterns observed in UMAP projections, which preserve non-linear manifold structure, may provide more meaningful insights into the network's actual generalization capabilities. The site- and modality-specific clustering is reduced when using UMAP projection, compared with PCA, yet it persists. This outcome underscores the critical importance of multi-center datasets and the necessity of data harmonization or domain adaptation techniques to accommodate diverse clinical contexts ~\cite{glocker_machine_2019}. A centered location of the OOD data for the first three PCA components should suggest that some robust CL or T1w-modality MS brain representations were learned despite the dominant presence of some sites in the Train set. 

\subsection{Limitations and future work}
An evident limitation in our study arises from the exclusive reliance on T1-weighted MRI modalities (MPRAGE and MP2RAGE). The persistent inability of models—and human observers—to reliably identify subpial lesions at 3T T1-weighted imaging confirms prior research emphasizing limited subpial lesion visibility at lower magnetic field strengths~\cite{beck_improved_2018}. Incorporating additional MRI contrasts (e.g., T2-weighted or fluid-suppressed sequences) or using ultra-high field data ~\cite{la_rosa_multiple_2022} might theoretically enhance segmentation quality. However, practical hurdles such as non-standardized acquisition protocols and limited availability across clinical sites significantly hinder this integration. This study again highlights that different CLs are visible on different MRI modalities, underscoring the importance of standardized lesion annotation criteria to facilitate consistent CL interpretation across studies and improve the clinical validity of automated segmentation tools.

Future studies should explore human-AI interaction systematically, assessing clinician trust in AI-generated segmentations and possible integration scenarios. Collecting feedback from the annotators about the missed and false-positive lesions would be essential for understanding the DL model decisions. 

From the AI-engineering perspective, there are several future directions aiming to improve in-domain performance and generalizability. We identified that varying sizes and per-subject loads, with overall small volumes of segmented instances, still represent a challenge for the proposed model, even in-domain (Figure ~\ref{fig:les_dist}). Potential remedies include multi-size labeling for improved small lesion segmentation~\cite{shang_segmenting_2024}, inclusion of nDSC to the loss for improving fairness across subjects~\cite{raina_tackling_2023, spagnolo_exploiting_2024}, or using pretrained transformer-CNN architectures (e.g., UNETR or SwinUnet) for adaptive context reasoning and semantic consistency properties. To improve the performance under the domain shift, one can adopt existing pretrained models for WML models~\cite{la_rosa_software_2020} or the nnU-Net framework~\cite{wald_revisiting_2025}, implement synthetic lesion generation strategies (e.g., CarveMix~\cite{zhang_carvemix_2023}), or consider test-time adaptation approaches~\cite{valanarasu_--fly_2024}. 

The aforementioned approaches would not resolve fundamental ambiguities inherent to distinguishing leukocortical from juxtacortical lesions, highlighting persistent challenges due to inherent lesion variability and high prevalence. Although quantifying exact cortical involvement remains problematic, future annotation protocols might benefit from marking uncertain lesions explicitly, enabling the use of soft labels and probabilistic approaches in training automated segmentation models.

\section{Conclusions}

This study benchmarks automated cortical lesion segmentation in multiple sclerosis using the nnU-Net framework. Our results show that a standard nnU-Net architecture, with task-specific adjustments like the Blob BCE + Dice Loss, effectively segments cortical lesions across different imaging centers and MRI protocols. Cross-site performance variability and the similarity between leukocortical and juxtacortical lesions remains challenging and necessitate clearer annotation guidelines, possibly involving soft labels. 

Automated segmentation tools demonstrated a potential for clinical workflows as supportive diagnostic or research aids. Their role could include providing automatic lesion segmentation as a second opinion to highlight lesions potentially missed by expert raters. Additionally, the lesion delineation procedure can be significantly simplified with the presence of automatic tools, even if manual corrections are required. 




\section*{Acknowledgment}

This work was supported by the Hasler Foundation Responsible AI program (MSxplain) and the Research Commission of the Faculty of Biology and Medicine (CRFBM) of UNIL. 
We acknowledge access to the facilities and expertise of the CIBM Center for Biomedical Imaging, a Swiss research center of excellence founded and supported by Lausanne University Hospital (CHUV), University of Lausanne (UNIL), École polytechnique fédérale de Lausanne (EPFL), University of Geneva (UNIGE), and Geneva University Hospitals (HUG).
AC is supported by EUROSTAR E!113682 HORIZON2020.
DR and CT are supported by the Intramural Research Program of NINDS, NIH. 
CVB received funding from the Fonds de Recherche Clinique (FRC) from Cliniques Universitaires Saint-Luc (CUSL). AS has the financial support of the Fédération Wallonie Bruxelles – FRIA du Fonds de la Recherche Scientifique – FNRS. SB is supported by the Funds Claire Fauconnier, Ginette Kryksztein \& José, and Marie Philippart-Hoffelt, managed by the King Baudouin Foundation. 
PM is supported by the Fondation Charcot Stichting Research Fund 2023, the Fund for Scientific Research (F.R.S, FNRS; grant \#40008331), Cliniques universitaires Saint-Luc “Fonds de Recherche Clinique”, and Biogen. 

\section*{Ethics statement}

Studies involving human data were approved by the local ethics committees; informed consent was obtained from all participants before study entry.

\appendix

\section{Bottleneck space analysis}
\label{ap:latent}

\begin{figure}[h!]
    \centering
    \includegraphics[width=0.6\linewidth]{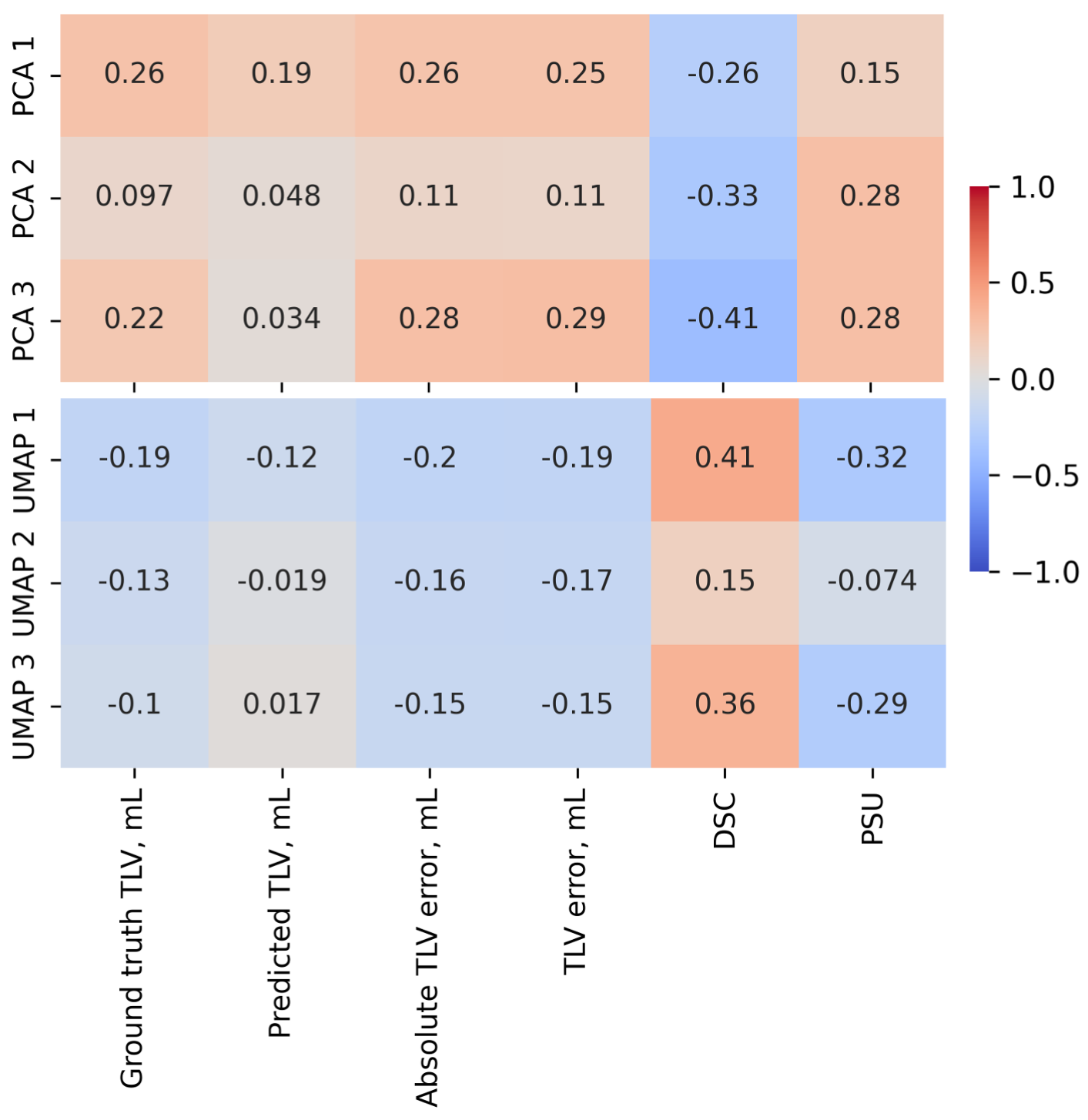}
    \caption{Pearson correlation between projected bottleneck features and continuous features.}
    \label{fig:latcont}
\end{figure}

The relationship between projected internal model features and continuous features (DSC, patient-wise model uncertainty, lesion volume errors) is quantified in Figure \ref{fig:latcont}. Mild correlation is observed between bottleneck feature components and continuous features, with the highest overall correlation values of DSC. 

\begin{figure}[h!]
    \centering
    \includegraphics[width=\linewidth]{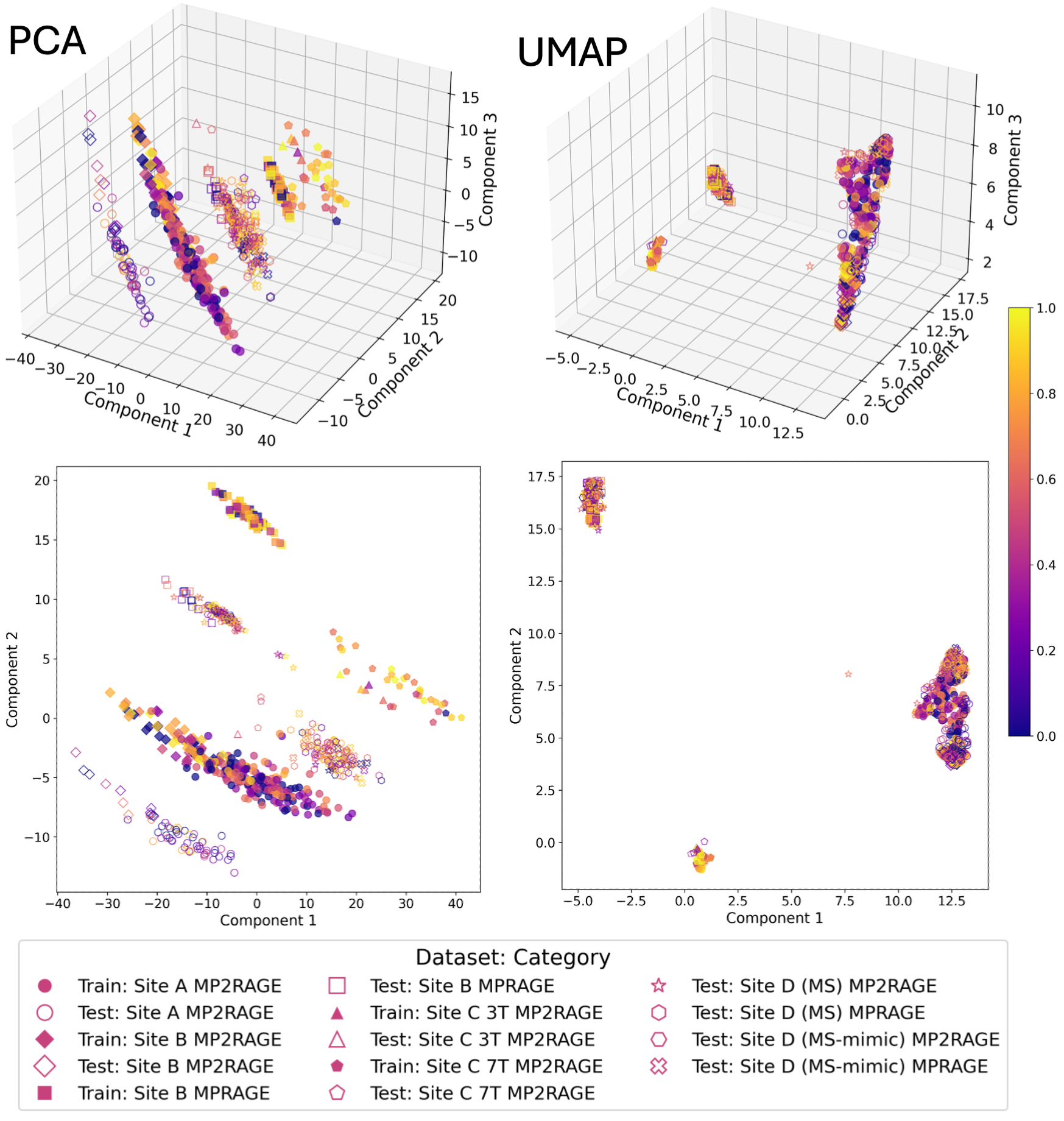}
    \caption{Distribution of patient-wise uncertainty ranking (computed with PSU measure) with respect to the internal model representations and sites. Latent features were projected with PCA (left) and UMAP (right). Original PSU values underwent the quantile transform to accommodate a skewed uncertainty distribution.}
    \label{fig:latunc}
\end{figure}

The patient structural uncertainty (PSU) quantifies the disagreement between individual ensemble predictions and the average ensemble prediction, reflecting both epistemic and aleatoric uncertainty (potential model errors or ambiguous data)~\cite{molchanova_structural-based_2025}. Figure \ref{fig:latunc} illustrates model features with uncertainty information. Low uncertainty examples are found across all PCA clusters of the bottleneck feature space, even in clusters distant from the Site A training set (e.g., Site B MPRAGE cluster). Conversely, Site C training and test examples consistently exhibit high uncertainty. This widespread higher uncertainty in training data across different sites may indicate significant data noise, hindering the learning of consistent representations.

\clearpage

\bibliographystyle{elsarticle-num-names} 
\bibliography{references-2}

\end{document}